\documentclass[11pt,a4paper]{article}
\usepackage{jcappub}

\def\gsim{\;\rlap{\lower 2.5pt
 \hbox{$\sim$}}\raise 1.5pt\hbox{$>$}\;}
\def\lsim{\;\rlap{\lower 2.5pt
   \hbox{$\sim$}}\raise 1.5pt\hbox{$<$}\;}
\def\ie{{\it i.e. }}
\def\eg{{\it e.g. }}

\title{Accurate Shear Measurement with Faint Sources}

\author{Jun Zhang$^{a,c}$, Wentao Luo$^b$, Sebastien Foucaud$^a$}
\affiliation{$^a$Center for Astronomy and Astrophysics, Department of Physics and Astronomy, Shanghai Jiao Tong University, 955 Jianchuan road, Shanghai, 200240, China} 
\affiliation{$^b$Key Laboratory for Research in Galaxies and Cosmology, Shanghai Astronomical Observatory, Nandan Road 80, Shanghai, 200030, China}
\affiliation{$^c$Texas Cosmology Center, the University of Texas at Austin, Austin, TX 78712, USA} 
\emailAdd{betajzhang@sjtu.edu.cn}

\abstract{For cosmic shear to become an accurate cosmological probe, systematic errors in the shear measurement method must be unambiguously identified and corrected for. Previous work of this series has demonstrated that cosmic shears can be measured accurately in Fourier space in the presence of background noise and finite pixel size, without assumptions on the morphologies of galaxy and PSF. The remaining major source of error is source Poisson noise, due to the finiteness of source photon number. This problem is particularly important for faint galaxies in space-based weak lensing measurements, and for ground-based images of short exposure times. In this work, we propose a simple and rigorous way of removing the shear bias from the source Poisson noise. Our noise treatment can be generalized for images made of multiple exposures through MultiDrizzle. This is demonstrated with the SDSS and COSMOS/ACS data. With a large ensemble of mock galaxy images of unrestricted morphologies, we show that our shear measurement method can achieve sub-percent level accuracy even for images of signal-to-noise ratio less than $5$ in general, making it the most promising technique for cosmic shear measurement in the ongoing and upcoming large scale galaxy surveys. }

\keywords{cosmology, large scale structure, gravitational lensing - methods, data analysis - techniques, image processing}

\arxivnumber{1312.5514}

\begin{document}
\maketitle
\flushbottom

\section{Introduction}
\label{intro}

Recent results from the cosmic microwave background observations have placed tight constraints on cosmological theories \citep{hinshaw13,planck}. The inflationary ${\rm \Lambda CDM}$ model remains to be the minimal best-fit model, though the physical origin of dark matter and dark energy is still unknown. Properties of the dark components can be further studied with large scale probes at late Universe, among which weak gravitational lensing is one of the few model-independent methods, as it is purely caused by gravity \citep{bs01,hj08,weinberg13}. On large scales, lensing induces weak but systematic shape distortions of background galaxies, whose statistical properties can be measured directly and compared to theoretical predictions \citep{kaiser00,van00,wittman00,massey07,mandelbaum13,van13}. Precise measurement of the weak lensing effect is one of the primary targets of a number of ongoing and planned large scale galaxy surveys (\eg, DES\footnote{www.darkenergysurvey.org}, Subaru HSC\footnote{www.naoj.org/Projects/HSC/}, Euclid\footnote{sci.esa.int/euclid/}, LSST\footnote{www.lsst.org}).

In contrast to its simple physical interpretation, weak lensing effect is difficult to measure for, but not limited to, the following several reasons: 1. the weak lensing signal (cosmic shear) is only of order a few percent, much less than the intrinsic dispersion of galaxy shapes; 2. the point spread function (PSF hereafter) alters the galaxy image at a level that is more significant than the lensing effect; 3. the pixelation effect due to the discrete nature of digital images can affect the shear measurement; 4. the background photon noise and the source Poisson noise can contaminate the shear signal; 5. the second order lensing effects should be taken into account \citep{jz11}. Among these problems, the PSF correction has been most extensively discussed \citep{kaiser95,lk97,hoekstra98,rrg00,kaiser2000,bridle01,bj02,rb03,hs03,mr05,kuijken06,miller07,nb07,kitching08,jz08}. Collaborative efforts have been made to compare the accuracies of different shear measurement methods under general conditions \citep{heymans06,massey2007,bridle10,kitching12,mandelbaum2013}. The goal is to achieve sub-percent level accuracy in shear measurement, so that the statistical power of the upcoming large scale weak lensing survey can be fully realized \citep{huterer06}.  

In the previous work of this series, a method based on the quadrupole moments of the galaxy power spectrum (PS hereafter) is developed \citep{jz08,jz10,jz11,zk11}. It is encouraging to note that in this method, most of the existing shear measurement problems have been solved in a model-independent way: the PSF correction does not require any assumptions about the morphologies of the galaxies or the PSF; the systematic error due to the background noise can be removed in a statistically rigorous way; the error due to the finite pixel size is negligible as long as it is smaller than about $1/3$ of the FWHM of the PSF; the method is accurate to the second order in shear/convergence. In addition, the method does not require a determination of the centroid of the galaxy, and the signal-to-noise ratio per galaxy can in principle be significantly increased due to the shape information from galaxy substructures. The overall image processing takes a very small amount of time (typically less than $10^{-2} {\rm CPU}\cdot {\rm second/galaxy}$), the majority of which is on the Fast Fourier Transformation of the image. Given these advantages, we are strongly motivated to solve the remaining problems in this method, the major one of which is the systematic error due to the source Poisson noise, as shown in \citet{jz11} (Z11 hereafter). This is the main purpose of this paper. In \S\ref{Z11}, we give a brief introduction of Z11.

The source Poisson noise is due to the finite exposure time. Faint galaxy images contain a very limited number of photons each, even though they may be selected as valid sources in low background cases, such as in space-based surveys, or by stacking many exposures in the ground-based observations. In the later case, the source Poisson noise is important if the shear measurement is carried out on individual exposure (\eg, CFHTLenS method described in \citep{miller13}). The faint galaxies may cause significant systematic errors in shear measurements. For example, it is shown in Z11 that the source Poisson noise can affect the shear measurement accuracy at percent level or more when the source photon number is less than $\sim 10^4$. This implies that for a telescope of about one meter in size, if the I-band data from galaxies of AB magnitude $25$ is used in the method of Z11, the overall exposure time on each source should be $\sim 3$ hours (one photon per second) to achieve sub-percent level accuracy in shear measurement. For sources fainter than 25 in magnitude, the exposure time should be higher correspondingly to satisfy the requirement.

Faint sources play a major role in shear measurement, as their number density is much larger than the brighter ones. However, the telescope exposure time is typically expensive. On the other hand, short exposure time is sometimes preferred for minimizing the effect of unstable observation conditions (\eg, PSF) and transient events (\eg, cosmic rays). We therefore have a strong motivation to understand that without sacrificing the accuracy in shear measurements, if the requirement on the source photon number is really necessary. We find that this goal is indeed achievable by extending the method of Z11, as shown in this work. 

The key idea is to realize that the PS of the Poisson noise is independent of the wave-number in Fourier space, unlike the PS of the source signal which is strongly suppressed by the PSF at large wave-numbers. The PS of the Poisson noise can therefore be estimated at large wave-numbers, and then subtracted at all wave-numbers to avoid shear bias in the method of Z11. The details of this idea are presented in \S\ref{single}\footnote{We note that the idea of estimating and subtracting the PS of the Poisson noise from the source PS is not new. Similar treatment has been widely applied in the studies of galaxy clustering \citep{thomas10,Ho12,kang12,anderson14} and CMB.}. For images combined from a number of exposures through, \eg, the MultiDrizzle algorithm \citep{fh02,koekemoer07}, we show that the PS of the Poisson noise of the image is a function of the wave-number, which is determined by the parameters of dithering and MultiDrizzle. In this case, the amplitude of the Poisson noise PS can still be estimated using Fourier modes at large wave-numbers, and subtracted at all wave-numbers according to the pre-determined form of the Poisson noise PS. By incorporating the above procedures into the method of Z11, we can remove the systematic shear bias due to the source Poisson noise in very general cases. This part of our work is shown in \S\ref{dithering}. In \S\ref{data}, we use the SDSS data and the COSMOS/ACS data to confirm our statements about the properties of the Poisson noise. 

We note that the noise bias in shear measurement was previously discussed by several authors \citep{refregier12,mv12,miller13,vio14} for methods that based on fitting the galaxy images with parameterized profiles. Their calculation and removal of the noise bias are therefore model-dependent. In contrast, the noise treatment proposed in this paper is free of assumptions on galaxy profile due to the model-independent nature of Z11. This feature is particularly important for shear measurement at high redshift, where the fraction of irregular galaxies is large \citep{bundy05}. In \S\ref{conclusions}, we give a conclusion of our method, as well as a brief discussion about some other existing shear recovery methods.

\section{Shear measurement in Fourier space}
\label{Z11}

Weak lensing stretches every background galaxy image along a certain direction. This effect is characterized by two parameters $(g_1, g_2)$, the so-called reduced shear, which are physically determined by the large scale structure along the line-of-sight. The purpose of weak lensing measurement is to find shear estimators as functionals of the observed galaxy image, the ensemble averages of which should recover the shear values accurately. The shear estimators can be used to measure either the spatial distribution or the correlation functions of the shear fields, making it possible to directly probe the dynamical properties of dark matter and dark energy on large scales. 

In the method of Z11, the shear estimators for the two components of the reduced shear $g_1$ and $g_2$ are defined based on the Fourier transform of the galaxy image. There are three components: $G_1$, $G_2$, and $N$, made of the multipole moments of the PS of the galaxy image in Fourier space:
\begin{eqnarray}
\label{shear_estimator}
G_1&=&-\frac{1}{2}\int d^2\vec{k}(k_x^2-k_y^2)T(\vec{k})M(\vec{k})\\ \nonumber
G_2&=&-\int d^2\vec{k}k_xk_yT(\vec{k})M(\vec{k})\\ \nonumber
N&=&\int d^2\vec{k}\left[k^2-\frac{\beta^2}{2}k^4\right]T(\vec{k})M(\vec{k})
\end{eqnarray}
where
\begin{equation}
T(\vec{k})=\left\vert\tilde{W}_{\beta}(\vec{k})\right\vert^2/\left\vert\tilde{W}_{PSF}(\vec{k})\right\vert^2,\;\;\; M(\vec{k})=\left\vert\tilde{f}^S(\vec{k})\right\vert^2-\left\vert\tilde{f}^B(\vec{k})\right\vert^2
\end{equation}
and $\tilde{f}^S(\vec{k})$ and $\tilde{f}^B(\vec{k})$ are the Fourier transformations of the galaxy image and a neighboring image of background noise respectively. Note that "neighboring" here simply means that the background noise image should be located close to the source image, so that they share similar background properies. Since the background noise image should have the same size as that of the source image, one can choose it from one of the eight immediate neighbors of the source image, avoiding those that contain identified sources. $\tilde{W}_{PSF}(\vec{k})$ is the Fourier transform of the PSF. $\tilde{W}_{\beta}(\vec{k})$ is the Fourier transform of an isotropic Gaussian function of scale radius $\beta$, which is defined as:
\begin{equation}
\label{beta}
W_{\beta}(\vec{x})=\frac{1}{2\pi\beta^2}\exp\left(-\frac{\left\vert\vec{x}\right\vert^2}{2\beta^2}\right).
\end{equation}
The factor $T(\vec{k})$ is used to convert the form of the PSF to the desired isotropic Gaussian function for correcting the PSF effect. The choice of $\beta$ should be somewhat larger than the scale radius of the original PSF $W_{PSF}$ to avoid singularities in the conversion, and should not be too large to cause significant information loss. In practice, we choose it to be about $20\%\sim 50\%$ larger than the radius of the original PSF. It is shown in Z11 that the ensemble averages of the shear estimators defined above do recover the shear values to the second order in accuracy (assuming that the intrinsic galaxy images are statistically isotropic), \ie, 
\begin{equation}
\label{shear_measure}
\frac{\left\langle G_1\right\rangle}{\left\langle N\right\rangle}=g_1+O(g_{1,2}^3),\;\;\;\frac{\left\langle G_2\right\rangle}{\left\langle N\right\rangle}=g_2+O(g_{1,2}^3)
\end{equation}
Note that the ensemble averages are taken for $G_1$, $G_2$, and $N$ separately. The details of how to use this type of shear estimators to measure the shear correlation functions are given in \citet{zk11}.

In practice, $G_1$, $G_2$, and $N$ are calculated using discrete Fourier transformation. Assuming the pixel size and the total number of pixels of a stamp image along the $x$ and $y$ directions are $(\Delta_1, N_1)$ and $(\Delta_2, N_2)$ respectively, the discrete Fourier transformation of the galaxy image is defined as:
\begin{equation}
\label{FT}
\tilde{f}(\vec{k}_j)=\sum_{i=1}^{N_T}f(\vec{x}_i)\exp\left[{\mathrm i}\vec{x}_i\cdot\vec{k}_j\right]
\end{equation}
where $i$ is the pixel index, and $j$ is the index for the discrete Fourier wave number, both of which are in the range of $[1,N_T]$, with $N_T=N_1\times N_2$. The components of $\vec{x}_i$ and $\vec{k}_j$ take the following values:
\begin{equation}
\label{kx}
\vec{x}_i=\left(m\Delta_1,n\Delta_2\right), \;\;\;\vec{k}_j=\left(2\pi u/(N_1\Delta_1), 2\pi v/(N_2\Delta_2)\right)
\end{equation}
with
\begin{eqnarray}
\label{mnuv}
&&m=0, 1, \cdots, N_1-1\\ \nonumber
&&n=0, 1, \cdots, N_2-1\\ \nonumber
&&u=-N_1/2+1,-N_1/2+2, \cdots, N_1/2\\ \nonumber
&&v=-N_2/2+1,-N_2/2+2, \cdots, N_2/2
\end{eqnarray}
Consequently, we have:
\begin{equation}
\label{abs_f}
\left\vert\tilde{f}(\vec{k}_j)\right\vert^2=\sum_{m=1}^{N_T}\sum_{n=1}^{N_T}f(\vec{x}_m)f(\vec{x}_n)\exp\left[{\mathrm i}(\vec{x}_m-\vec{x}_n)\cdot\vec{k}_j\right]
\end{equation}
The shear estimators can be written as:
\begin{eqnarray}
\label{shear_estimator_dis}
G_1&=&-\frac{C}{2}\sum_{j=1}^{N_T}\left[(\vec{k}_j)_x^2-(\vec{k}_j)_y^2\right]T(\vec{k}_j)M(\vec{k}_j)\\ \nonumber
G_2&=&-C\sum_{j=1}^{N_T}(\vec{k}_j)_x(\vec{k}_j)_yT(\vec{k}_j)M(\vec{k}_j)\\ \nonumber
N&=&C\sum_{j=1}^{N_T}\left[\left\vert\vec{k}_j\right\vert^2-\frac{\beta^2}{2}\left\vert\vec{k}_j\right\vert^4\right]T(\vec{k}_j)M(\vec{k}_j)
\end{eqnarray}
with $C=4\pi^2/(N_T\Delta_1\Delta_2)$, and
\begin{equation}
\label{shear_estimator_dis_para1}
T(\vec{k}_j)=\left\vert\tilde{W}_{\beta}(\vec{k}_j)\right\vert^2/\left\vert\tilde{W}_{PSF}(\vec{k}_j)\right\vert^2
\end{equation}
\begin{equation}
\label{shear_estimator_dis_para2}
M(\vec{k}_j)=\left\vert\tilde{f}^S(\vec{k}_j)\right\vert^2-\left\vert\tilde{f}^B(\vec{k}_j)\right\vert^2
\end{equation}

The validity of the master formulae, including eq.(\ref{shear_measure},\ref{shear_estimator_dis}) and the definitions in eq.(\ref{shear_estimator_dis_para1},\ref{shear_estimator_dis_para2}), has been tested extensively in the discrete case in our previous work. The remaining problem is on the shear measurement error caused by the source Poisson noise, which refers to the random counting error of the source photons. The discrete formulation of the shear estimators is the preparation for our discussion about how to remove the counting noise in the rest of the paper.

\section{Extension of Z11 for single-exposure images}
\label{single}

The shear estimators defined in eq.(\ref{shear_measure},\ref{shear_estimator_dis},\ref{shear_estimator_dis_para1},\ref{shear_estimator_dis_para2}) can be significantly biased if the source image does not contain enough photons, as shown in Z11. The fluctuation of the photon number distribution across the galaxy scale can alter the shape of the galaxy, and thereby affect the shear measurement accuracy. This effect is different from that by the background noise, as discussed in Z11. It is found that to achieve sub-percent level accuracy in shear recovery, one needs to collect at least about $10^4$ photons per galaxy. In this section, we demonstrate that the bound on the source photon number can be alleviated. With both analytic reasoning and numerical examples, we show how to remove the shear bias due to the source Poisson noise by a slight modification of the forms of the shear estimators. The discussion in this section is for images made of single exposures. More general cases are considered in \S\ref{dithering}. 

\subsection{The algorithm}
\label{algorithm1}

Our idea is to utilize the fact that statistically, the PS of the Poisson noise is independent of the wave-number. Its amplitude can be estimated at the large wave-number limit, at which the PS of the source signal is sub-dominant due to filtering by the PSF. The estimated Poisson noise PS can then be subtracted from the PS of the image at all wave-numbers. This operation is particularly suitable in the method of Z11, as its shear estimators are all linear functions of the PS. The same procedure should be repeated on the neighboring image of background noise, as the Poisson noise in the source image is partly due to the background photons.

In general, let us define the readout of the $i^{th}$ pixel of an image as $f_i=\bar{f}_i+\Delta f_i$, where $\Delta f_i$ is the Poisson noise, and $\bar{f}_i$ is the statistical expectation of the signal. Note that $\Delta f_i$ here only refers to the Poisson noise (due to both the source and the background), not the physical fluctuation of the background flux, which is contained in the term $\bar{f}_i$. Assuming each pixel readout has a linear response to the number of photons it receives, we have $f_i=a N_i+b$, where $N_i$ is the number of photons received by the $i^{th}$ pixel, and $a$, $b$ are the coefficients of the linear relation. Note that despite the complication of the raw image processing pipeline, the method proposed in this paper only requires the linear response assumption to hold locally, \ie, the spatial variation of the coefficients $a$ and $b$ over the scale of the stamp image of the galaxy is small. Consequently, $\Delta f_i$ obeys the Poisson statistics: 
\begin{equation}
\langle\Delta f_i \Delta f_j\rangle=a^2\langle\Delta N_i\Delta N_j\rangle=a^2\delta_{ij}\bar{N}_i=a (\bar{f}_i-b)\delta_{ij}
\end{equation}
where $\delta_{ij}$ is the Kronecker delta function. The amplitudes of the Fourier modes can be written as:
\begin{equation}
\label{abs_f_df}
\left\vert\tilde{f}_j\right\vert^2=\sum_{m=1}^{N_T}\sum_{n=1}^{N_T}(\bar{f}_m+\Delta f_m)(\bar{f}_n+\Delta f_n)\exp\left[{\mathrm i}(\vec{x}_m-\vec{x}_n)\cdot\vec{k}_j\right]
\end{equation}
Taking the ensemble average of eq.(\ref{abs_f_df}), we get: 
\begin{equation}
\label{abs_f_df_ensem}
\left\langle\left\vert\tilde{f}_j\right\vert^2\right\rangle=\left\langle\left\vert\tilde{\bar{f}}_j\right\vert^2\right\rangle+\left\langle F\right\rangle
\end{equation}
with $F=a\sum_{m=1}^{N_T}(\bar{f}_m-b)$. Clearly, the net effect of the Poisson noise is to raise the PS on all scales by the same amount. On the other hand, the PS of the physical signal (including both the source and background) is strongly suppressed by the PSF at large wave-numbers. The way of removing the Poisson noise effect is straightforward: one can estimate the PS of the Poisson noise at large enough wave-numbers, and then subtract it on all scales. A critical wave-number $k_c$ can be defined for this purpose. The choice of $k_c$ can be quite flexible. For example, it can be close to the Nyquist spatial-frequency/wave-number of either direction. For a given $k_c$, we have:
\begin{equation}
\label{abs_f_df_ensem2}
\left.\left\langle\left\vert\tilde{f}_j\right\vert^2\right\rangle\right\vert_{\vert\vec{k}_j\vert>k_c}\approx\left\langle F\right\rangle
\end{equation}
so, for each image, $F$ can be estimated as:
\begin{equation}
\label{abs_f_df_ensem3}
F\approx\frac{\sum_{\vert\vec{k}_j\vert>k_c}\left\vert\tilde{f}_j\right\vert^2}{\sum_{\vert\vec{k}_j\vert>k_c}}
\end{equation}
where $\sum_{\vert\vec{k}_j\vert>k_c}$ in the denominator refers to the number of modes satisfying $\vert\vec{k}_j\vert>k_c$. The Poisson noise contamination to the PS can be removed as:
\begin{equation}
\label{abs_f_df_ensem4}
\left\vert\tilde{\bar{f}}_j\right\vert^2\approx\left\vert\tilde{f}_j\right\vert^2-F
\end{equation}
Following these thoughts, to remove the source Poisson noise effect, we only need to modify the definition of $M(\vec{k})$ in eq.(\ref{shear_estimator_dis_para2}) as: 
\begin{equation}
\label{shear_estimator_dis_para3}
M(\vec{k})=\left\vert\tilde{f}^S(\vec{k})\right\vert^2-F^S-\left\vert\tilde{f}^B(\vec{k})\right\vert^2+F^B
\end{equation}
with
\begin{equation}
\label{shear_estimator_dis_para4}
F^S=\frac{\sum_{\vert\vec{k}_j\vert>k_c}\left\vert\tilde{f}^S(\vec{k}_j)\right\vert^2}{\sum_{\vert\vec{k}_j\vert>k_c}}, \;\;\; F^B=\frac{\sum_{\vert\vec{k}_j\vert>k_c}\left\vert\tilde{f}^B(\vec{k}_j)\right\vert^2}{\sum_{\vert\vec{k}_j\vert>k_c}}
\end{equation}

Eq.(\ref{shear_estimator_dis_para3}) is an extension of eq.(\ref{shear_estimator_dis_para2}). The two additional terms $F^S$ and $F^B$ are estimates of the PS of the Poisson noise on the source image and background image respectively. Note that $F^S$ contains contributions from the Poisson noise of both the galaxy and the background. The difference between $F^S$ and $F^B$ therefore yields a statistically unbiased estimate of the PS due to the source Poisson noise only, which is subtracted in eq.(\ref{shear_estimator_dis_para3}) for shear measurement accuracy.

The shear estimators defined by eq.(\ref{shear_measure},\ref{shear_estimator_dis},\ref{shear_estimator_dis_para1},\ref{shear_estimator_dis_para3},\ref{shear_estimator_dis_para4}) are unbiased in the presence of both the source Poisson noise and the background noise for single-exposure images. This is  tested with a large ensemble of simulated galaxy images in the next section.

\subsection{Numerical test}
\label{ne}

We adopt the simulation pipeline of Z11. Every simulated galaxy image is placed at the center of a $64\times 64$ grid, each cell of which is regarded as a CCD pixel, whose size is used as the length unit in the rest of the paper. Without loss of generality, we adopt the truncated Moffat profile for the form of the PSF (used in the GREAT08 project):
\begin{equation}
\label{PSFs}
W_{PSF}(r)\propto\left[1+\left(\frac{r}{r_d}\right)^2\right]^{-3.5}{\mathrm H}(r_c-r)
\end{equation}
where ${\mathrm H}(r_c-r)$ is the Heaviside step function. We set $r_c=3r_d$ and $r_d=3$ in our simulation. Note that this arrangement sets the pixel size to be about $1/3$ of the FWHM of the PSF, which is shown in Z11 to be the maximum pixel size allowed for avoiding significant pixelation effect. Each galaxy is made of a collection of point sources of equal luminosity, the positions of which are determined by the end points of 40 random walk steps. Every step has a random size between $0$ and $1$, pointing at a random direction. To limit the size of the galaxy, the random walks restart from the grid center if its distance to the grid center is more than $10$. Once generated, the point sources are displaced according to the input shear, and then assigned to the pixels according to the PSF. Note that in shear measurement, we transform the PSF to the isotropic Gaussian form defined in eq.(\ref{beta}) with $\beta=0.7r_d$.  

Two types of photon noise are added onto the galaxy images, including the background noise and the source Poisson noise. The background noise includes the physical fluctuation of the background flux and its Poisson fluctuation on the CCD. For simplicity, the background noise in our simulation is added as a Poisson noise of a uniform background flux. Both types of Poisson noise are generated as Gaussian random variables. We use Gaussian random variable to mimic Poisson noise in this paper for two reasons: 1. The 1-point PDF (Probability Distribution Function) of Poisson fluctuation approaches that of Gaussian fluctuation in the limit of large photon number in each pixel (say, 100); 2. More importantly, removal of the source Poisson noise effect only relies on the fact that the PS of the Poisson noise is scale-independent, not on the exact form of its 1-point PDF.

\begin{figure}
\centerline{\epsfxsize=9cm\epsffile{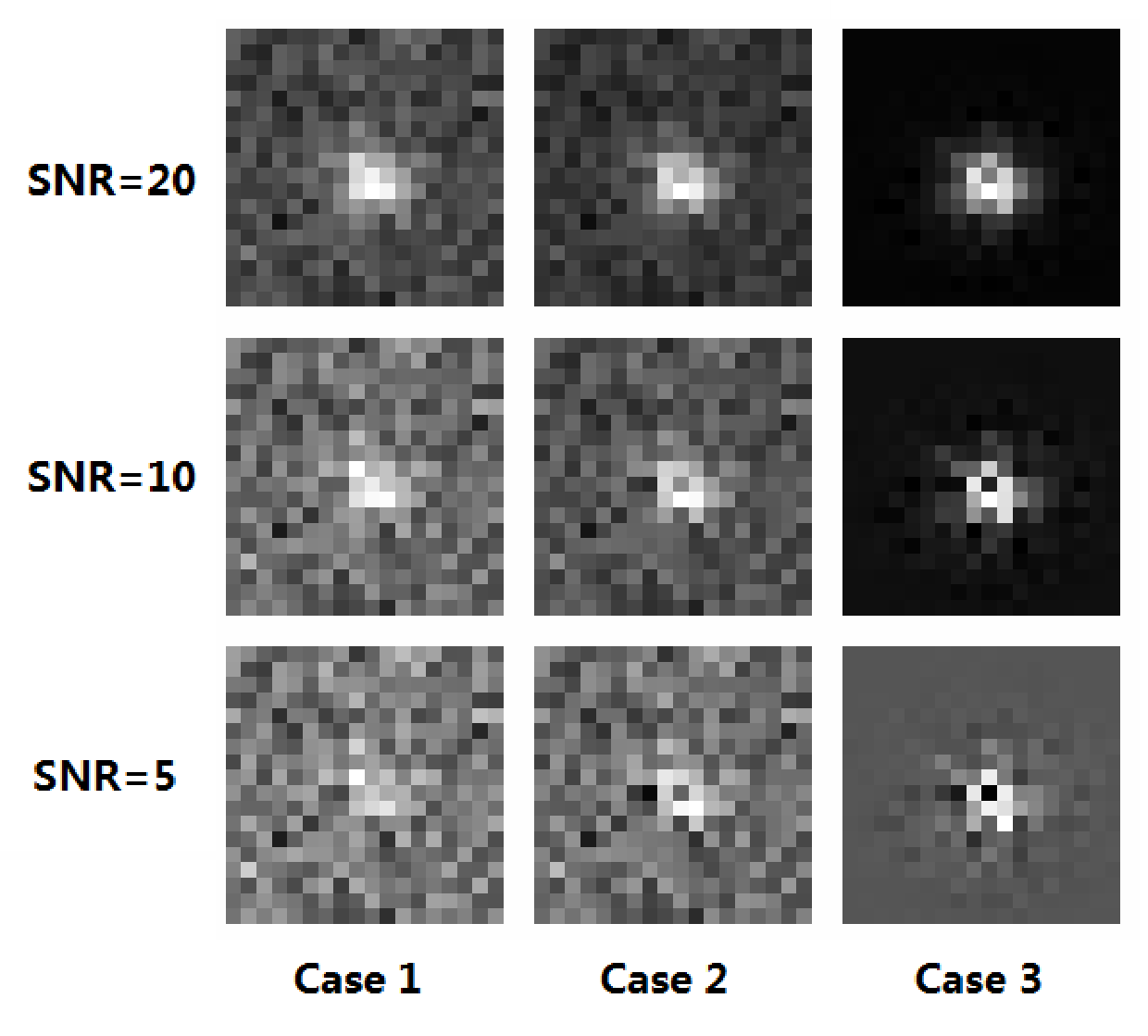}}
\caption{Examples of galaxy images generated with the pipeline of \S\ref{ne}.}
\label{no_dither}
\end{figure}

Since the background noise and source Poisson noise play different roles in shear measurement, we may define the signal-to-noise ratio separately as ${\rm SNR}_B$ (for the background noise) and ${\rm SNR}_S$ (for the source Poisson noise). Since these two types of noise are not correlated, the total ${\rm SNR}$ is:
\begin{equation}
\label{SNR}
\frac{1}{{\rm SNR}^2}=\frac{1}{{\rm SNR}_B^2}+\frac{1}{{\rm SNR}_S^2}
\end{equation}
For any given ${\rm SNR}$, we test the shear recovery accuracy in three cases: ${\rm SNR}_B/{\rm SNR}_S=0$, $1$, and $\infty$, which are denoted below as case $1$, $2$, and $3$ respectively. Sample images are shown in fig.\ref{no_dither}. Note that for each galaxy image, we generate a companion image of background noise with a different random seed in the test. 

As usual, we quantify the shear recovery accuracy with the multiplicative and additive bias defined as:
\begin{equation}
\label{shear_error}
g_{1,2}^{measured}=(1+m_{1,2})g_{1,2}^{input}+c_{1,2}
\end{equation}
where the subscripts $1$, $2$ refer to the first and second components of the cosmic shear respectively. We use six sets of input shear values $(g_1,g_2)$: (0.05,-0.05), (0.03,-0.03), (0.01,-0.01), (-0.01,0.01), (-0.03,0.03), and (-0.05,0.05). The multiplicative bias $m$ and additive bias $c$ are computed by fitting the results of all 6 sets to the linear relation in eq.(\ref{shear_error}). This numerical experiment is repeated for different values of ${\rm SNR}_B$ and ${\rm SNR}_S$, which are realized by only adjusting the amplitudes of the source Poisson noise and the background noise, without changing the random seeds. This arrangement makes it easy and efficient to observe the effect of our noise treatment. To reduce the shape noise, every four galaxies in our simulations are generated by a single galaxy rotated by $n\times 45^{\circ}$ ($n=0,1,2,3$).

In table \ref{result_regular1_no_correction}, we show the shear recovery accuracy of the original Z11 using $2\times 10^7$ mock galaxies. It is clear that the source Poisson noise, which is present in case 2 and 3, can lead to a significant multiplicative bias. In comparison, table \ref{result_regular1} shows the accuracy of the slightly modified version of Z11 defined in eq.(\ref{shear_measure},\ref{shear_estimator_dis},\ref{shear_estimator_dis_para1},\ref{shear_estimator_dis_para3},\ref{shear_estimator_dis_para4}) with $3\times 10^8$ mock galaxies. It demonstrates that the modified Z11 can indeed recover shear at a sub-percent level accuracy in the presence of both the background and the source Poisson noise for galaxies with the overall ${\rm SNR}$ as low as 5. The success likely continues at even lower ${\rm SNR}$, though identification of sources becomes increasingly difficult. In applying the modified Z11, we set the critical wave-number $k_c$ defined in eq.(\ref{shear_estimator_dis_para4}) to be $3/4$ of the Nyquist spatial-frequency/wave-number of one direction. 

\begin{table}
\centering
\begin{tabular}{cccc}

\hline\hline		
SNR                     &  Case 1                 & Case 2              &   Case 3             \\
\hline\hline
 & $(m_1,m_2)(10^{-3})$ &  &\\
        & $(c_1,c_2)(10^{-5})$ & &\\
\hline		                                                             
$20$ & $(1.7\pm 0.8, -0.7\pm 0.8)$     & $(15.9\pm 0.5, 14.1\pm 0.5)$    & $(30.2\pm 0.2, 29.9\pm 0.2) $ \\
        & $(-4.9\pm 2.7, 1.5\pm 2.7)$      & $(-3.1\pm 1.8, 0.9\pm 1.8)$    & $(-0.1\pm 0.8, 0.3\pm 0.8)$ \\
\hline
$10$ & $(5.6\pm 2.5, -0.4\pm 2.5)$     & $(65.1\pm 1.5, 60.8\pm 1.5)$    & $(131.7\pm 0.5, 131.2\pm 0.5)$ \\
         & $(-14.7\pm 8.4, 5.8\pm 8.4)$     & $(-8.2\pm 5.0, 3.2\pm 5.0)$    & $(-0.2\pm 1.7, 0.6\pm 1.7)$ \\
\hline
$5$ & $(20.2\pm 9.0 , 3.1\pm 9.0)$    & $(316.0\pm 6.1, 302.3\pm 6.1)$   & $(864.4\pm 1.7, 862.8\pm 1.7)$ \\
        & $(-43\pm 31, 23\pm 31)$    & $(-28\pm 21, 14\pm 21)$   & $(-0.2\pm 5.7, 2.5\pm 5.7)$ \\
\hline\hline		
\end{tabular}
\caption{The multiplicative bias $m_{1,2}$ and the additive bias $c_{1,2}$ of the reduced shear measured using the original Z11 method. In each data cell, the first row shows the multiplicative bias $(m_1, m_2)$ in unit of $10^{-3}$, and the second row shows the additive bias $(c_1, c_2)$ in unit of $10^{-5}$. The results are shown for three values of the overall SNR: 20,10,5. For each SNR, three cases are considered: 1. only background noise; 2. the background noise and source Poisson noise have equal amplitudes; 3. only source Poisson noise.}
\label{result_regular1_no_correction}
\end{table}

\begin{table}
\centering
\begin{tabular}{cccc}

\hline\hline		
SNR                     &  Case 1                 & Case 2              &   Case 3             \\
\hline\hline
 & $(m_1,m_2)(10^{-3})$ &  &\\
        & $(c_1,c_2)(10^{-5})$ & &\\
\hline		                                                             
$20$ & $(-0.2\pm 0.2, -0.2\pm 0.2)$     & $(-0.2\pm 0.1, -0.3\pm 0.1)$    & $(-0.08\pm 0.05, -0.3\pm 0.05) $ \\
        & $(-1.0\pm 0.7, 1.4\pm 0.7)$      & $(-0.6\pm 0.4, 1.1\pm 0.4)$    & $(0.0\pm 0.2, 0.3\pm 0.2)$ \\
\hline
$10$ & $(-0.6\pm 0.6, -0.3\pm 0.6)$     & $(-0.4\pm 0.4, -0.4\pm 0.4)$    & $(-0.1\pm 0.1, -0.4\pm 0.1)$ \\
         & $(-2.6\pm 2.1, 3.7\pm 2.1)$     & $(-1.6\pm 1.2, 2.6\pm 1.2)$    & $(0.0\pm 0.4, 0.6\pm 0.4)$ \\
\hline
$5$ & $(-2.2\pm 2.3, -0.7\pm 2.3)$    & $(-1.1\pm 1.2, -0.9\pm 1.2)$   & $(0.0\pm 0.2, -0.6\pm 0.2)$ \\
        & $(-7.9\pm 7.9, 10.6\pm 7.9)$    & $(-4.6\pm 4.1, 6.6\pm 4.1)$   & $(0.1\pm 0.8, 1.2\pm 0.8)$ \\
\hline\hline		
\end{tabular}
\caption{Similar to table \ref{result_regular1_no_correction}, except that the shear measurement method is a slightly modified version of Z11 defined in eq.(\ref{shear_measure},\ref{shear_estimator_dis},\ref{shear_estimator_dis_para1},\ref{shear_estimator_dis_para3},\ref{shear_estimator_dis_para4}).}
\label{result_regular1}
\end{table}

\section{Extension of Z11 for multi-exposure images}
\label{dithering}

\subsection{Dithering \& MultiDrizzle}
\label{background}

In Z11, the ideal observing condition is that the CCD pixel size is not larger than a third of the FWHM of the PSF, \ie, the galaxy images should be reasonably sampled. This condition is however not always satisfied, particularly for the space-based projects that are diffraction-limited. For example, in the COSMOS survey, the CCD pixel size of the Advanced Camera for Surveys (ACS) Wide-Field Channel (WFC) detector is 15$\mu$mm, corresponding to $\sim 0.05"$/pixel, while the FWHM of the PSF by the HST optics in the F814W filter is about $0.085"$, or $\sim 0.1"$ if taken into account the convolution by the detector pixel \citep{koekemoer07}. A natural solution for the under-sampling problem is to take multiple exposures at slightly shifted directions (by sub-pixel amounts), a technique which is termed dithering \citep{fh02}. 

Note that taking multiple exposures of the same patch of the sky also provides a way of removing transient impacts such as cosmic rays. This could be achieved by down-weighting the impacted pixels when combining the exposures, as done in the COSMOS survey with MultiDrizzle. More generally, for a given observing condition, it is worth studying how to optimally divide the total observing time (including the telescope overhead time) into a number of exposures for the purpose of weak lensing, the conclusion of which does not seem trivial. For example, more exposures allow more options in treating the cosmic ray impacts, but each exposure becomes more noisy. The conclusion of such a study will likely depend on the implied shear measurement method, and many other factors such as the stability of the telescope optics. This topic is beyond the scope of this work. Here, we focus on how dithering affects the behavior of the Poisson noise, especially the source Poisson noise, as shown below.

If the dithers are placed very well, one can simply interlace the pixels from different exposures onto a finer grid to improve sampling. In this case, the Poisson noise of the finer pixels are not correlated, and the noise treatment defined in \S\ref{algorithm1} is valid. Interlacing remains as an option for image reconstruction with multiple dithers, though it is usually considered not feasible due to small positioning errors and non-uniform shifts across the pixels. Currently, the most popular image reconstruction technique is called drizzle, which is widely used for HST images. 

\subsection{The algorithm}
\label{multidrizzle}

The purpose of drizzle is to construct images of finer pixel sizes from multiple exposures of dithered locations. The value of an output pixel is calculated as a weighted sum of the neighboring input pixels of different exposures. Because of this, drizzle introduces correlated noise between pixels, and consequently, the PS of the source Poisson noise has a non-flat shape in Fourier space that is determined by the drizzle strategy. Following the idea of \S\ref{algorithm1}, one can estimate the amplitude of the Poisson noise PS at large enough wave-numbers, and then remove it at all wave-numbers. More specifically, let us present the idea using the COSMOS data as an example. 

Let us denote the output image as $g(\vec{x}_i)$, with $\vec{x}_i$ being the position of the $i^{th}$ pixel, and the dithered exposures as $f^{\alpha}(\vec{x}^{\alpha}_i)$, with $\alpha$ and $i$ being the image and pixel indices respectively. Note that here, the vectors $\vec{x}_i$ and $\vec{x}^{\alpha}_i$ refer to the actual position angles in the sky, not the relative positions within the stamp images. In COSMOS, since each output image is constructed from four input images, we have $\alpha=1, 2, 3, 4$. The output image is then a weighted sum of the input images:
\begin{equation}
\label{drizzle}
g(\vec{x}_i)=\sum_{\alpha=1}^{4}\sum_jW_{ij}^{\alpha}C(\vec{x}_i-\vec{x}^{\alpha}_j)f^{\alpha}(\vec{x}^{\alpha}_j)
\end{equation}
where the kernel function $C$ characterizes how the flux of each pixel on the input images should be assigned to its neighboring pixels on the output image, and $W_{ij}^{\alpha}$ is the additional weighting. 

The kernel can generally be written as a function of the relative position of the output pixel with respect to the input pixel. It could be a square top-hat function or a Gaussian function \citep{rhodes07}. In the former case, the function $C(\vec{x}_i-\vec{x}^{\alpha}_j)$ is simply proportional to the overlapping area between the input pixel located at $\vec{x}^{\alpha}_j$ and the output pixel at $\vec{x}_i$. If the kernel is a Gaussian function, it means that $C$ is proportional to the integral of a circular Gaussian function (centered at the center of the input pixel) over the area of the output pixel (square shape). The FWHM of the Gaussian kernel is determined by the parameter ${\rm pixfrac}$ in unit of the input pixel size (see details in \S\ref{ntest2}). 

The weighting function $W_{ij}^{\alpha}$ is affected by the presence of cosmic rays, chip defects, chip gaps, etc. A careful study of $W_{ij}^{\alpha}$ requires knowledge in hardware design and the scheduling of the telescope time, which are specific for individual surveys. Here, we simply set $W_{ij}^{\alpha}=1$, which is indeed possible if the number of exposures for each patch of the sky could be increased. The effect of uneven weighting is discussed in \S\ref{data}. Consequently, the Poisson noises of the input and output images are related through:
\begin{equation}
\label{drizzle1}
\Delta g_i=\sum_{\alpha=1}^{4}\sum_jC(\vec{x}_i-\vec{x}^{\alpha}_j)\Delta f^{\alpha}_j
\end{equation}
where we have simplified the notations of $g(\vec{x}_i)$ and $f^{\alpha}(\vec{x}^{\alpha}_j)$ without confusions. The correlation function of the Poisson noise in the output image is:
\begin{equation}
\label{drizzle2}
\langle \Delta g_i\Delta g_j\rangle=\sum_{\alpha=1}^{4}\sum_{\beta=1}^{4}\sum_l\sum_mC(\vec{x}_i-\vec{x}^{\alpha}_l)C(\vec{x}_j-\vec{x}^{\beta}_m)\langle \Delta f^{\alpha}_l\Delta f^{\beta}_m\rangle
\end{equation}
Using the fact that $\langle \Delta f^{\alpha}_l\Delta f^{\beta}_m\rangle=\langle (\Delta f^{\alpha}_l)^2\rangle\delta_{\alpha\beta}\delta_{lm}$, we get:
\begin{equation}
\label{drizzle3}
\langle \Delta g_i\Delta g_j\rangle=\sum_{\alpha=1}^{4}\sum_l C(\vec{x}_i-\vec{x}^{\alpha}_l)C(\vec{x}_j-\vec{x}^{\alpha}_l)\langle (\Delta f^{\alpha}_l)^2\rangle
\end{equation}
which shows that drizzle introduces correlation of the Poisson noise between different output pixels. The PS of the Poisson noise is therefore given by:
\begin{eqnarray}
\label{drizzle4}
\left\langle\Delta\left\vert\tilde{g}(\vec{k})\right\vert^2\right\rangle&=&\sum_{ij}\langle \Delta g_i\Delta g_j\rangle\exp\left[{\mathrm i}(\vec{x}_i-\vec{x}_j)\cdot\vec{k}\right]\\ \nonumber
&=&\sum_{\alpha=1}^{4}\sum_l \langle (\Delta f^{\alpha}_l)^2\rangle\sum_{ij}C(\vec{x}_i-\vec{x}^{\alpha}_l)C(\vec{x}_j-\vec{x}^{\alpha}_l)\exp\left[{\mathrm i}(\vec{x}_i-\vec{x}_j)\cdot\vec{k}\right]
\end{eqnarray}
Note that eq.(\ref{drizzle4}) contains contributions from the Poisson noises of both the source photons and the background, \ie, 
\begin{eqnarray}
\label{drizzle5}
\left\langle\Delta\left\vert\tilde{g}(\vec{k})\right\vert^2\right\rangle&=&\left\langle\Delta^S\left\vert\tilde{g}(\vec{k})\right\vert^2\right\rangle+\left\langle\Delta^B\left\vert\tilde{g}(\vec{k})\right\vert^2\right\rangle\\ \nonumber
&=&\sum_{\alpha=1}^{4}\sum_l \left[\langle (\Delta^S f^{\alpha}_l)^2\rangle+\langle (\Delta^B f^{\alpha}_l)^2\rangle\right]\\ \nonumber
&\times&\sum_{ij}C(\vec{x}_i-\vec{x}^{\alpha}_l)C(\vec{x}_j-\vec{x}^{\alpha}_l)\exp\left[{\mathrm i}(\vec{x}_i-\vec{x}_j)\cdot\vec{k}\right]
\end{eqnarray}
with $\Delta^S$ and $\Delta^B$ symbolizing the Poisson noises of the source and background respectively. Since the kernel $C$ generally has a very small width (comparable to the pixel size), the summation over $\vec{x}_i$ and $\vec{x}_j$ in eq.(\ref{drizzle5}) yields the discrete Fourier transformation of the kernel $C$ as long as $\vec{x}^{\alpha}_l$ is not too close to the edges of the output image. This leads to a subtle difference in the spectral shapes of Poisson noises of the source and the background. The source photons are concentrated only in the central region of each stamp image, therefore for the source Poisson noise, the last line of eq.(\ref{drizzle5}) is simply given by the Fourier transform of the kernel $C$, \ie, 
\begin{eqnarray}
\label{drizzle6}
\left\langle\Delta^S\left\vert\tilde{g}(\vec{k})\right\vert^2\right\rangle&=&\sum_{\alpha=1}^{4}\sum_l \langle (\Delta^S f^{\alpha}_l)^2\rangle\sum_{ij}C(\vec{x}_i-\vec{x}^{\alpha}_l)C(\vec{x}_j-\vec{x}^{\alpha}_l)\exp\left[{\mathrm i}(\vec{x}_i-\vec{x}_j)\cdot\vec{k}\right]\\ \nonumber
&=&\left\vert\tilde{C}(\vec{k})\right\vert^2\sum_{\alpha=1}^{4}\sum_l \langle (\Delta^S f^{\alpha}_l)^2\rangle.
\end{eqnarray}
The background Poisson noise $\Delta^Bf^{\alpha}_l$ is present everywhere, therefore in eq.(\ref{drizzle5}), the summations over $\vec{x}_i$ or $\vec{x}_j$ contributed by the background Poisson noise do not exactly yield discrete Fourier transformations of the kernel $C$, \ie,
\begin{eqnarray}
\label{drizzle7}
\left\langle\Delta^B\left\vert\tilde{g}(\vec{k})\right\vert^2\right\rangle&=&\sum_{\alpha=1}^{4}\sum_l \langle (\Delta^B f^{\alpha}_l)^2\rangle\sum_{ij}C(\vec{x}_i-\vec{x}^{\alpha}_l)C(\vec{x}_j-\vec{x}^{\alpha}_l)\exp\left[{\mathrm i}(\vec{x}_i-\vec{x}_j)\cdot\vec{k}\right]\\ \nonumber
&=&\sum_{\alpha=1}^{4}\sum_l \langle (\Delta^B f^{\alpha}_l)^2\rangle \left\vert\tilde{\mathcal{C}}^{\alpha}_l(\vec{k})\right\vert^2
\end{eqnarray}
As a result, in Fourier space, the shape of $\langle\Delta^B\left\vert\tilde{g}(\vec{k})\right\vert^2\rangle$ and $\langle\Delta^S\left\vert\tilde{g}(\vec{k})\right\vert^2\rangle$ are slightly different. 

For our purpose, we should use $\tilde{C}(\vec{k})$ as the shape of the PS of the source Poisson noise. Eq.(\ref{drizzle6}) shows that the $\tilde{C}(\vec{k})$ has a non-trivial shape in Fourier space, determined by the parameters in the dithering strategy and MultiDrizzle. Following the idea that the amplitude of the Poisson noise PS can be estimated at large wave-numbers, and then removed on all scales according to the pre-determined spectral shape, we can remove the source Poisson noise effect for multi-exposure images by modifying eq.(\ref{shear_estimator_dis_para3},\ref{shear_estimator_dis_para4}) in the definitions of our shear estimators as:
\begin{equation}
\label{shear_estimator_dis_para5}
M(\vec{k})=\left\vert\tilde{f}^S(\vec{k})\right\vert^2-F^S\left\vert\tilde{C}(\vec{k})\right\vert^2-\left\vert\tilde{f}^B(\vec{k})\right\vert^2+F^B\left\vert\tilde{C}(\vec{k})\right\vert^2
\end{equation}
\begin{equation}
\label{shear_estimator_dis_para6}
F^S=\frac{\sum_{\vert\vec{k}_j\vert>k_c}\left\vert\tilde{f}^S(\vec{k}_j)\right\vert^2}{\sum_{\vert\vec{k}_j\vert>k_c}\left\vert\tilde{C}(\vec{k}_j)\right\vert^2},\;\;\; F^B=\frac{\sum_{\vert\vec{k}_j\vert>k_c}\left\vert\tilde{f}^B(\vec{k}_j)\right\vert^2}{\sum_{\vert\vec{k}_j\vert>k_c}\left\vert\tilde{C}(\vec{k}_j)\right\vert^2}
\end{equation}
Note that although the PS of the background Poisson noise does not have exactly the same shape as $\left\vert\tilde{C}(\vec{k})\right\vert^2$, its contribution to $F^S$ is statistically canceled by $F^B$, the counterpart of $F^S$ derived from a neighboring image of background noise. The net result of this operation is therefore equivalent to the removal of the source Poisson noise. In summary, to avoid the shear bias due to the source Poisson noise for images combined through MultiDrizzle, our shear estimators take the forms defined in eq.(\ref{shear_measure},\ref{shear_estimator_dis},\ref{shear_estimator_dis_para1},\ref{shear_estimator_dis_para5},\ref{shear_estimator_dis_para6}).

\subsection{Numerical test}
\label{ntest2}

Our pipeline for numerical test in this section is designed so that the output images mimic those from the COSMOS data. It consists of the following two steps:

1. For every galaxy image, we generate four exposures of the same ${\rm SNR}_S$ and ${\rm SNR}_B$. The position of each exposure is slightly shifted from the other ones by a half of the pixel size along the x-axis, or the y-axis, or both directions of the grid, forming a square dither pattern.    

2. The four exposures are combined using the MultiDrizzle algorithm with a Gaussian and isotropic convolution kernel, with ${\rm scale}=0.6$ and ${\rm pixfrac}=0.8$.

In the first step, each exposure is built on a $64\times 64$ grid. Every galaxy is made of a collection of point sources, the positions of which are determined by random walks as described in \S\ref{ne}. We still use the Moffat profile as the form of the PSF, and we choose $r_d=2$ instead of $3$ to let the exposures be under-sampled. Note that our ratio of the FWHM of the PSF to the pixel size is similar to that of the COSMOS/ACS raw image data. 

\begin{figure}
\centerline{\epsfxsize=9cm\epsffile{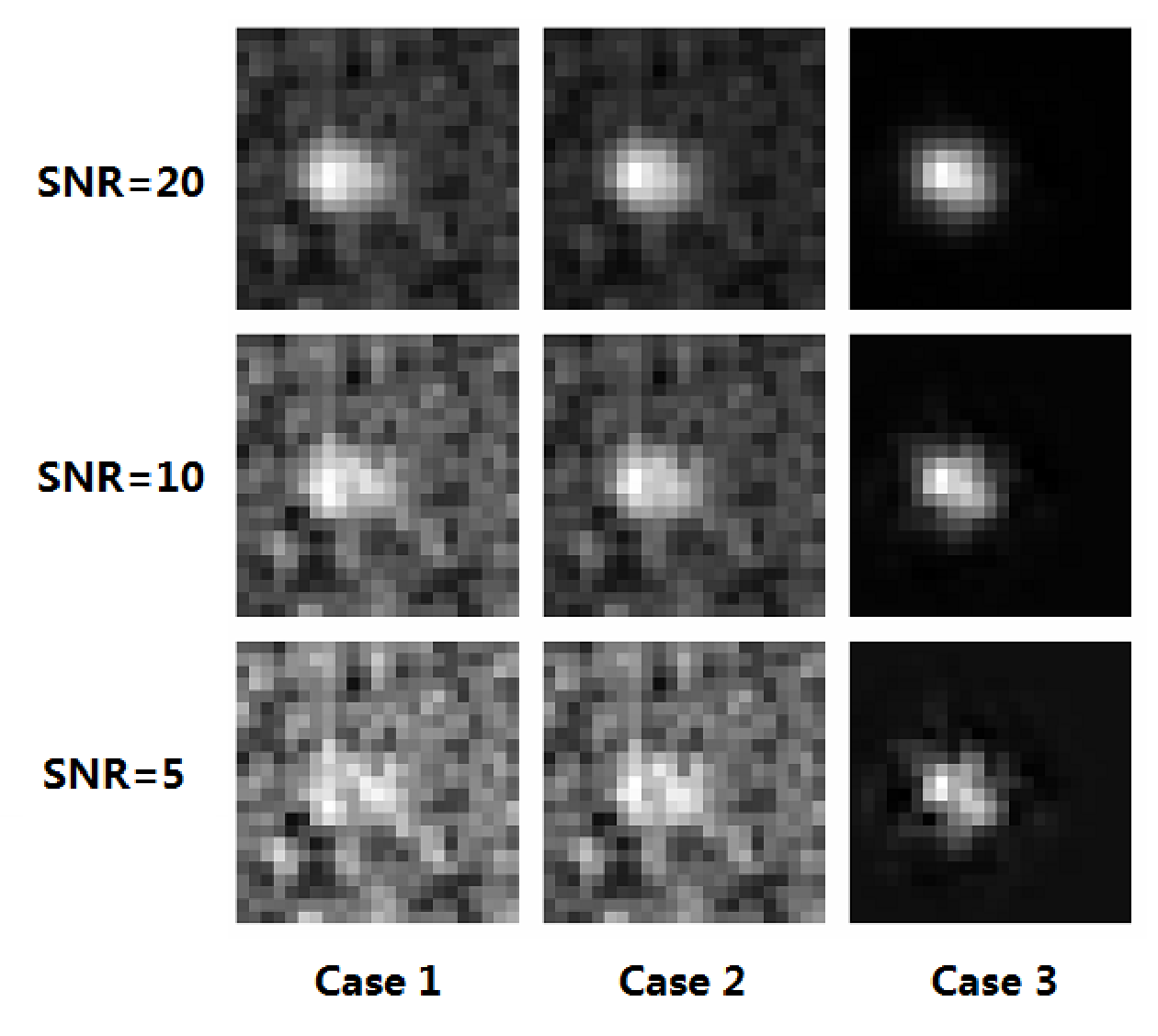}}
\caption{Examples of galaxy images generated with MultiDrizzle. The details are described in \S\ref{ntest2}.}
\label{with_dither}
\end{figure}

In the second step, the pixel size of the output image is $60\%$ of the input pixel size. This is specified by the parameter "${\rm scale}$". In the case of Gaussian convolution kernel, the parameter "${\rm pixfrac}$" determines the FWHM of the kernel in unit of the input pixel size. We also rotate the output image frame by about $10$ degrees with respect to the input image frame, although such a rotation does not exist in the lensing study of COSMOS \citep{rhodes07}. We find that our conclusion is not affected by the presence of the rotation of the output frame. The output grid is $64\times 64$, and centered close to the centers of the input images. The PSF in the output image is still a Moffat profile, with $r_d=2/0.6(\approx 3.33)$ in unit of the output pixel size. Note again that in shear recovery, we transform the PSF to the isotropic Gaussian form defined in eq.(\ref{beta}) with $\beta=0.7r_d$. Fig.\ref{with_dither} shows examples of the output images of different ${\rm SNR}$ in three cases.  

\begin{figure}
\centerline{\epsfxsize=9cm\epsffile{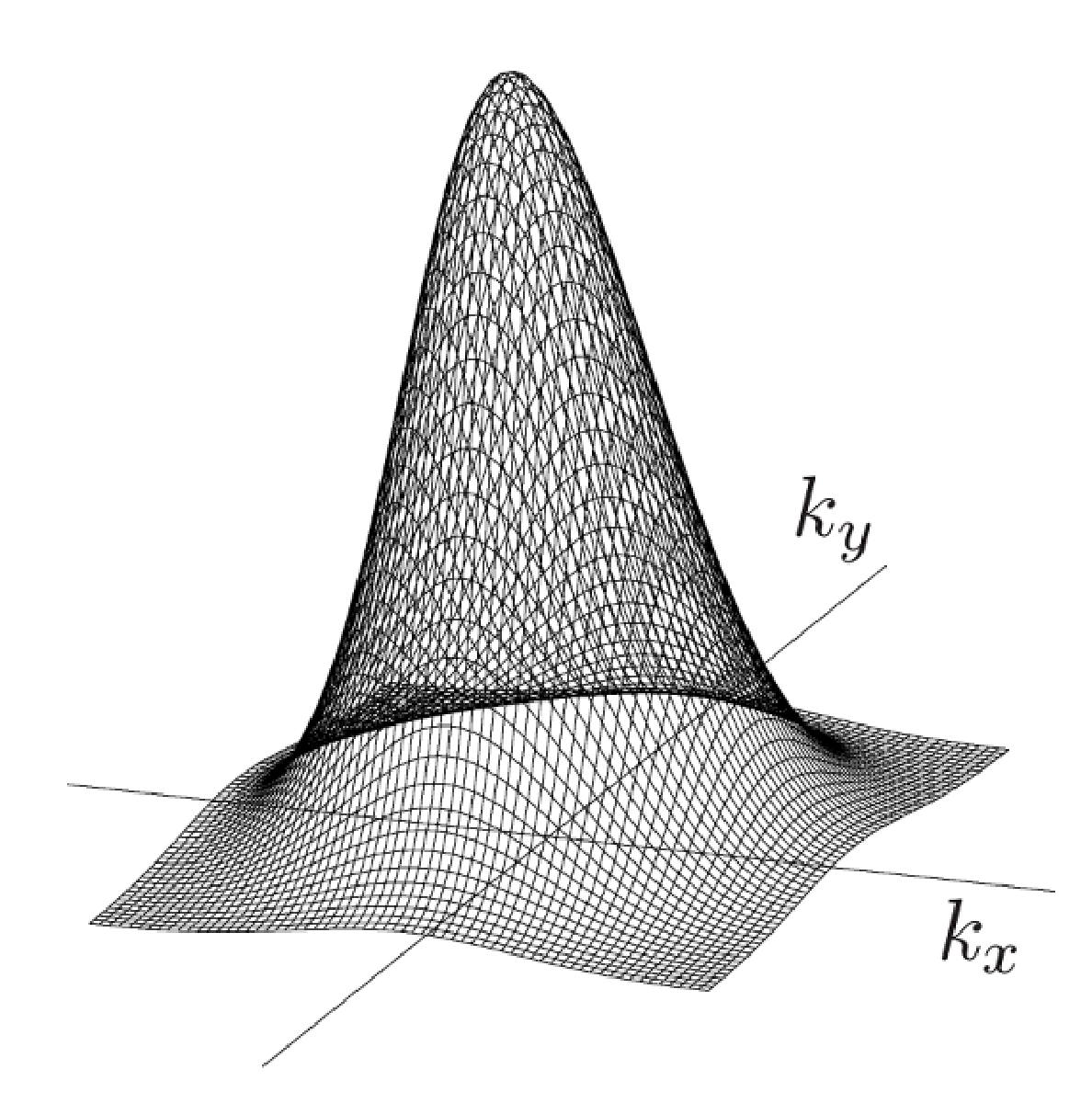}}
\caption{The shape of the PS of the source poisson noise $\left\vert\tilde{C}(\vec{k})\right\vert^2$ for the MultiDrizzle parameters described in \S\ref{ntest2}.}
\label{poisson_power}
\end{figure}

We compute the Poisson noise PS shape $\left\vert\tilde{C}(\vec{k})\right\vert^2$ using the parameters of MultiDrizzle, according to its definition in eq.(\ref{drizzle6}). It can also be measured numerically to any desired accuracy using a large number of simulated galaxy images processed by MultiDrizzle. Note that in doing so, the ambient background noise should not be included due to the boundary effect, as we discussed in \S\ref{multidrizzle}. Fig.\ref{poisson_power} shows $\left\vert\tilde{C}(\vec{k})\right\vert^2$ for the MultiDrizzle parameters described in this section.

To quantify the shear recovery accuracy, we use the same set of shear values as those given in \S\ref{ne}. Every four galaxy images are from the same galaxy rotated by integer multiples of $45^{\circ}$ to minimize the shape noise. Different values of ${\rm SNR}_B$ and ${\rm SNR}_S$ are realized by only adjusting the amplitudes of the background noise and the source Poisson noise, without changing the random seeds. 

Similar to \S\ref{ne}, we show in table \ref{result_dither1_no_correction} the results of shear recovery of the original Z11 method with $2\times 10^7$ mock galaxies. It is clear that the source Poisson noise can cause a significant multiplicative bias. In comparison, the modified version of Z11 defined in eq.(\ref{shear_measure},\ref{shear_estimator_dis},\ref{shear_estimator_dis_para1},\ref{shear_estimator_dis_para5},\ref{shear_estimator_dis_para6}) (with $k_c$ being $3/4$ of the Nyquist spatial-frequency/wave-number along one direction of the grid) can remove such a multiplicative bias to a very high accuracy. This is shown in table \ref{result_dither1} with $4\times 10^8$ mock galaxies. 

\begin{table}
\centering
\begin{tabular}{cccc}

\hline\hline		
SNR                 &  Case 1                 & Case 2              &   Case 3             \\
\hline\hline		                                                             
 & $(m_1, m_2)(10^{-3})$    &    & \\
         & $(c_1, c_2)(10^{-5})$    &    &  \\
\hline
$20$ & $(1.1\pm 0.6, 0.8\pm 0.6)$    & $(16.4\pm 0.4, 16.2\pm 0.4)$    & $(32.3\pm 0.2, 32.1\pm 0.2)$ \\
         & $(-1.0\pm 1.9, -4.3\pm 1.9)$    & $(-0.4\pm 1.3, -4.4\pm 1.3)$    & $(1.0\pm 0.6, -4.0\pm 0.6)$ \\
\hline
$10$ & $(2.0\pm 1.7, 0.5\pm 1.7)$     & $(65.8\pm 1.0, 65.0\pm 1.0)$    & $(138.5\pm 0.4, 138.0\pm 0.4)$ \\
         & $(-3.8\pm 5.7, -3.0\pm 5.6)$     & $(-2.2\pm 3.5, -4.4\pm 3.5)$    & $(1.7\pm 1.4, -4.6\pm 1.4)$ \\
\hline
$5$ & $(6.1\pm 6.0, -0.7\pm 5.9)$    & $(321.7\pm 4.2, 318.1\pm 4.1)$   & $(931.5\pm 1.5, 930.3\pm 1.5)$ \\
        & $(-12\pm 20, 5\pm 20)$    & $(-9\pm 14, -1\pm 14)$   & $(6.6\pm 5.0, -8.4\pm 5.0)$ \\
\hline\hline		

\end{tabular}
\caption{The multiplicative bias $m_{1,2}$ and the additive bias $c_{1,2}$ of the reduced shear measured using the original Z11 method. The images are generated from multiple exposures through MultiDrizzle, as described in \S\ref{ntest2}. In each data cell, the first row shows the multiplicative bias $(m_1, m_2)$ in unit of $10^{-3}$, and the second row shows the additive bias $(c_1, c_2)$ in unit of $10^{-5}$. The results are shown for three values of the overall SNR: 20,10,5. For each SNR, three cases are considered: 1. only background noise; 2. the background noise and source Poisson noise have equal amplitudes; 3. only source Poisson noise. }
\label{result_dither1_no_correction}
\end{table}

\begin{table}
\centering
\begin{tabular}{cccc}

\hline\hline		
SNR                 &  Case 1                 & Case 2              &   Case 3             \\
\hline\hline		                                                             
 & $(m_1, m_2)(10^{-3})$    &    & \\
         & $(c_1, c_2)(10^{-5})$    &    &  \\
\hline
$20$ & $(0.9\pm 0.1, 1.1\pm 0.1)$    & $(0.9\pm 0.08, 1.0\pm 0.08)$    & $(0.9\pm 0.04, 1.0\pm 0.05)$ \\
         & $(0.5\pm 0.4, -2.9\pm 0.4)$    & $(0.6\pm 0.3, -2.9\pm 0.3)$    & $(0.9\pm 0.1, -2.8\pm 0.1)$ \\
\hline
$10$ & $(0.8\pm 0.4, 1.3\pm 0.4)$     & $(0.9\pm 0.2, 1.2\pm 0.2)$    & $(0.9\pm 0.08, 0.9\pm 0.08)$ \\
         & $(0.6\pm 1.3, -2.7\pm 1.3)$     & $(0.5\pm 0.7, -2.8\pm 0.7)$    & $(0.8\pm 0.3, -2.7\pm 0.3)$ \\
\hline
$5$ & $(0.7\pm 1.3, 1.8\pm 1.3)$    & $(0.8\pm 0.7, 1.5\pm 0.7)$   & $(0.8\pm 0.2, 0.9\pm 0.2)$ \\
        & $(2.1\pm 4.6, -1.9\pm 4.5)$    & $(0.8\pm 2.4, -2.6\pm 2.4)$   & $(0.5\pm 0.6, -2.6\pm 0.6)$ \\
\hline\hline		

\end{tabular}
\caption{Similar to table \ref{result_dither1_no_correction}, except that the shear recovery method is the slightly modified version of Z11 defined in eq.(\ref{shear_measure},\ref{shear_estimator_dis},\ref{shear_estimator_dis_para1},\ref{shear_estimator_dis_para5},\ref{shear_estimator_dis_para6}). }
\label{result_dither1}
\end{table}

\section{Comparison with real data}
\label{data}

The PS of the source Poisson noise $\left\vert\tilde{C}(\vec{k})\right\vert^2$ is the key function in our method. For images made of single exposures,  $\left\vert\tilde{C}(\vec{k})\right\vert^2$ is simply a constant. We use the SDSS data to demonstrate it. The function $\left\vert\tilde{C}(\vec{k})\right\vert^2$ has a nontrivial form for images combined from multiple exposures through MultiDrizzle. This is shown with the COSMOS data.

\subsection{The SDSS data}
\label{SDSS}

We use the SDSS data as examples for images made of single exposures. We randomly select about 20000
sources flagged as galaxies from the 7th data release of Sloan Digital Sky Survey \citep{sdss7}, with r-band apparent magnitude in the range between 14 and 21. Each source is contained in a $50\times 50$ postage-stamp image, which is extracted from fpC file centered on the source. We further remove images with unpleasant features, e.g., those with bright edges (mostly due to a neighboring source), or with saturated pixels. 

\begin{figure}
\centerline{\epsfxsize=9cm\epsffile{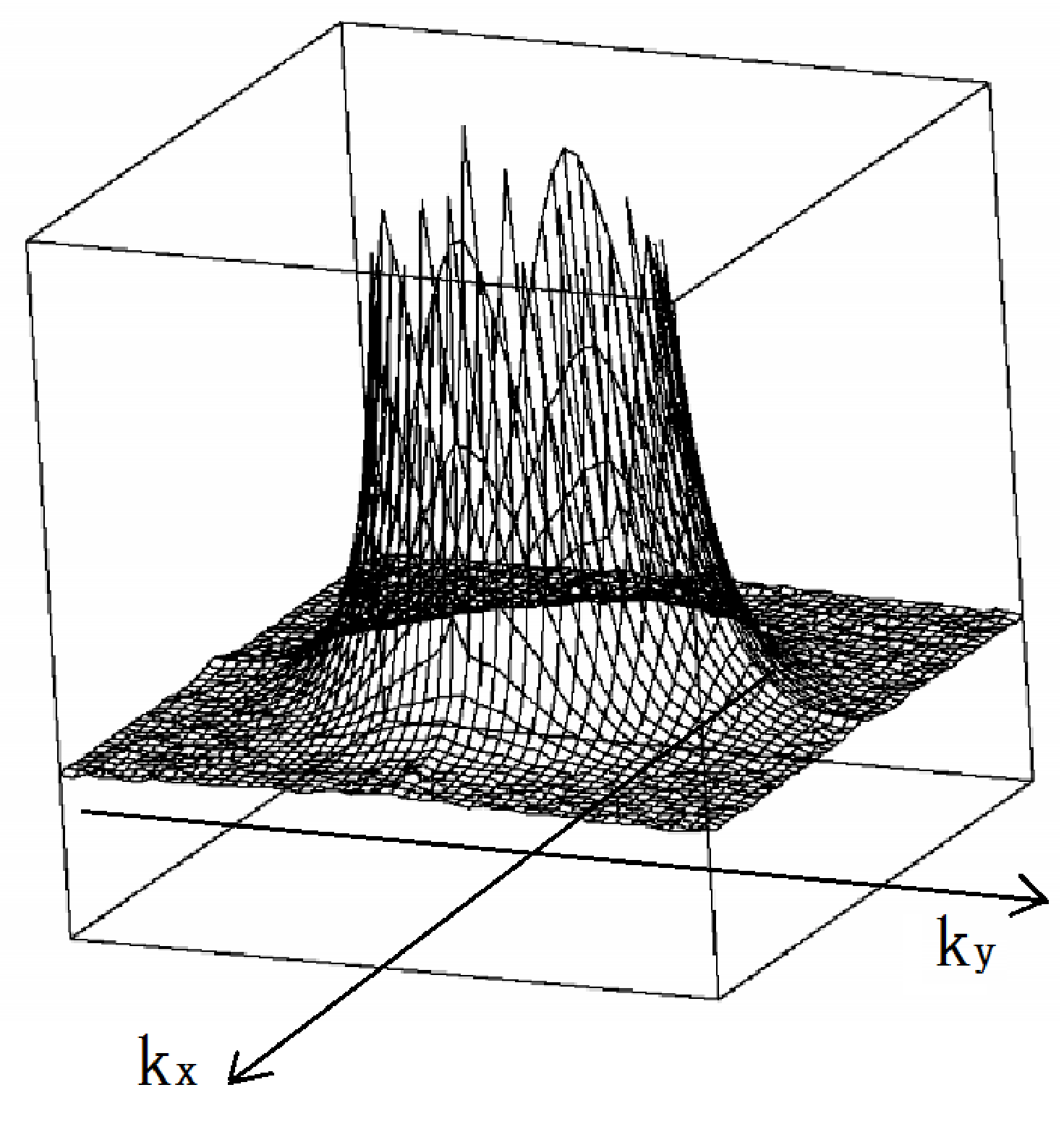}}
\caption{The stacked PS of the SDSS images ($50\times 50$ each). The four sides on the bottom of the plot are the directions of $k_x$ and $k_y$ respectively. The vertical axis shows the amplitude of the PS, with zero defined at the bottom of the cube. The central region of the cube corresponds to the spectrum at small wave-numbers, and is therefore cut out due to the dominance of the source spectrum.}
\label{poisson_power_sdss}
\end{figure}

To visualize the form of the Poisson noise PS, one can stack the PS of many different source images. Fig.\ref{poisson_power_sdss} shows such stacked PS with almost all source images available (excluding the brightest one percent of the sources because some of them are so bright that they may dominate the stacked PS). In the figure, the four sides on the bottom of the plot are the directions of $k_x$ and $k_y$ respectively. The vertical axis shows the amplitude of the PS, with zero defined at the bottom of the cube. The central region of the cube corresponds to the spectrum at small wave numbers, and is therefore cut out due to the dominance of the source spectrum. The figure clearly shows that at large wave numbers, the stacked PS of the images is dominated by the power of the Poisson noise, which is indeed flat (independent of $k$). 

\begin{figure}
\centerline{\epsfxsize=9cm\epsffile{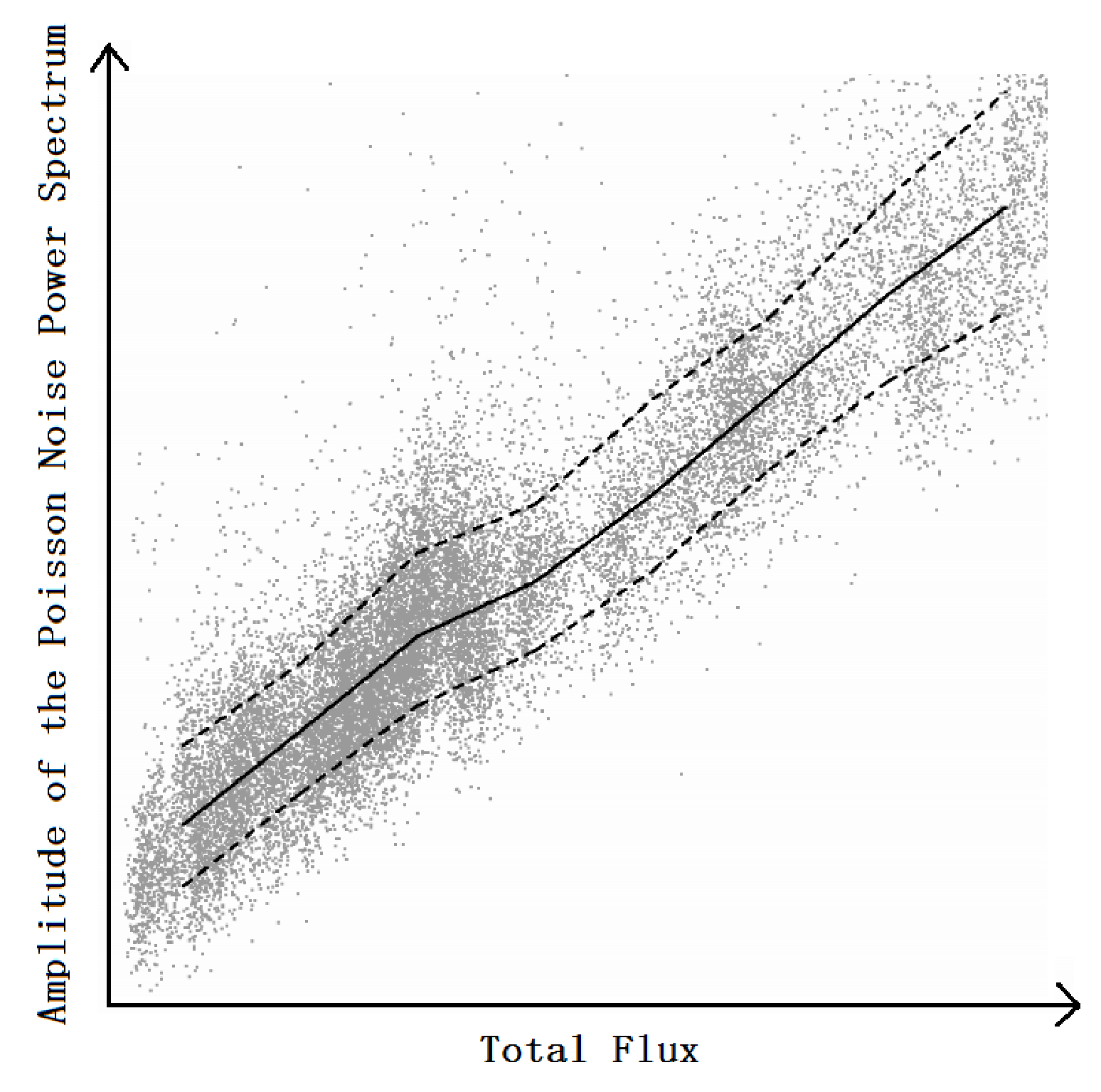}}
\caption{The relation between the Poisson noise PS and the total flux of the SDSS source image. The amplitude of the Poisson noise PS is measured by averaging the power of the Fourier transform of the source image at wave-number larger than 0.75 times the Nyquist spatial-frequency along either side of the image. The data are divided into 8 bins equally spaced in terms of the total flux. The solid curve connects the median values of the Poisson noise PS in neighboring bins, and 68 percent of the points are within the region enclosed by the dotted lines in each bin. Both axes are in linear scales. Since the absolute scales are not important in this study, they are not marked.}
\label{power_flux_sdss}
\end{figure}

The PS of Poisson noise in fig.\ref{poisson_power_sdss} contains the contributions from both the source and the background. To see the role of the source Poisson noise, in fig.\ref{power_flux_sdss}, we plot the amplitude of the Poisson noise PS against the total flux of the source stamp for all the selected galaxies. Note that the span of the total flux in the plot is mostly due to the range of the source luminosity, not due to the fluctuation of the background. The amplitude of the Poisson noise PS is measured by averaging the power of the Fourier transform of the source image at wave-number larger than 0.75 times the Nyquist spatial-frequency along either side of the image. Fig.\ref{power_flux_sdss} shows that the PS of the Poisson noise is linearly dependent on the total flux, agreeing with the theoretical expectation of eq.(\ref{abs_f_df_ensem}). In the figure, the data are divided into eight bins equally spaced in terms of the total flux. The solid curve connects the median values of the Poisson noise PS in neighboring bins, and 68 percent of the points are within the region enclosed by the dotted lines in each bin.

\subsection{The COSMOS/ACS data}
\label{COSMOS}

For images made of multiple exposures, we use the COSMOS/ACS data as an example to show the form of the PS of the source Poisson noise $\left\vert\tilde{C}(\vec{k})\right\vert^2$. This is done by taking the difference between the PS of the source image and that of a neighboring image of background noise, \ie, we use:
\begin{equation}
\label{measure_C}
\left\vert\tilde{C}(\vec{k})\right\vert^2\propto\left\langle\left\vert\tilde{f}^S(\vec{k})\right\vert^2-\left\vert\tilde{f}^B(\vec{k})\right\vert^2\right\rangle \quad {\rm if}\quad \vert\vec{k}\vert >k_c
\end{equation}
Note that the shape of $\left\vert\tilde{C}(\vec{k})\right\vert^2$ is not measurable at small wave-numbers due to the presence of the source power. 

We use version 2.0 of the "unrotated" ACS/WFC data, which is reduced for the purpose of lensing \citep{koekemoer07,leauthaud07,rhodes07} with the charge transfer efficiency problem corrected \citep{massey10}. The images were taken in the "Broad I" band through the wide F814W filter. The construction of the source catalog is described in \citet{leauthaud07}. Both the image data and the catalog are publicly available through the Infrared Science Archive (IRSA) database\footnote{http://irsa.ipac.caltech.edu/Missions/cosmos.html}.

The source catalog contains about $1.2\times 10^6$ sources\footnote{In the image file named "acs\_I\_100108+0213\_unrot\_sci\_20.fits", we identify a feature across the middle of the image, similar to the shape of a chip gap. For this reason, this image is not included in our analysis.}, all of which can in principle be used for visualizing the form of $\left\vert\tilde{C}(\vec{k})\right\vert^2$ at large wave-numbers. Note that stars and galaxies are equally good for this purpose. In practice, we apply the following requirements in source selection: 1. We use sources with MAG\_AUTO $\ge 26$, so that the stacked PS of the source Poisson noise is visible in a larger region in Fourier space;
2. The source image should be contained in a $64\times 64$ grid, and the boundaries of the source (defined in the catalog) should be at least six pixels away from the boundaries of the image;
3. In both the source image and the background noise image, each pixel within 6-pixel distance to the boundaries should be dimmer than 4 times the rms of the background noise (estimated locally). 

The last requirement is to avoid large aliasing powers due to bright sources located near the edges of the images. Note that the insignificant aliasing powers due to the unavoidable presence of unidentified/faint sources near the edges of the source image and the neighboring background noise image should statistically cancel out each other in eq.(\ref{measure_C}). The image of background noise is chosen among the eight immediate neighbors of the source image. From those (neighbors) that satisfy the last requirement above, we choose the one whose maximal pixel readout is the least. 

\begin{figure}
\centerline{\epsfxsize=14cm\epsffile{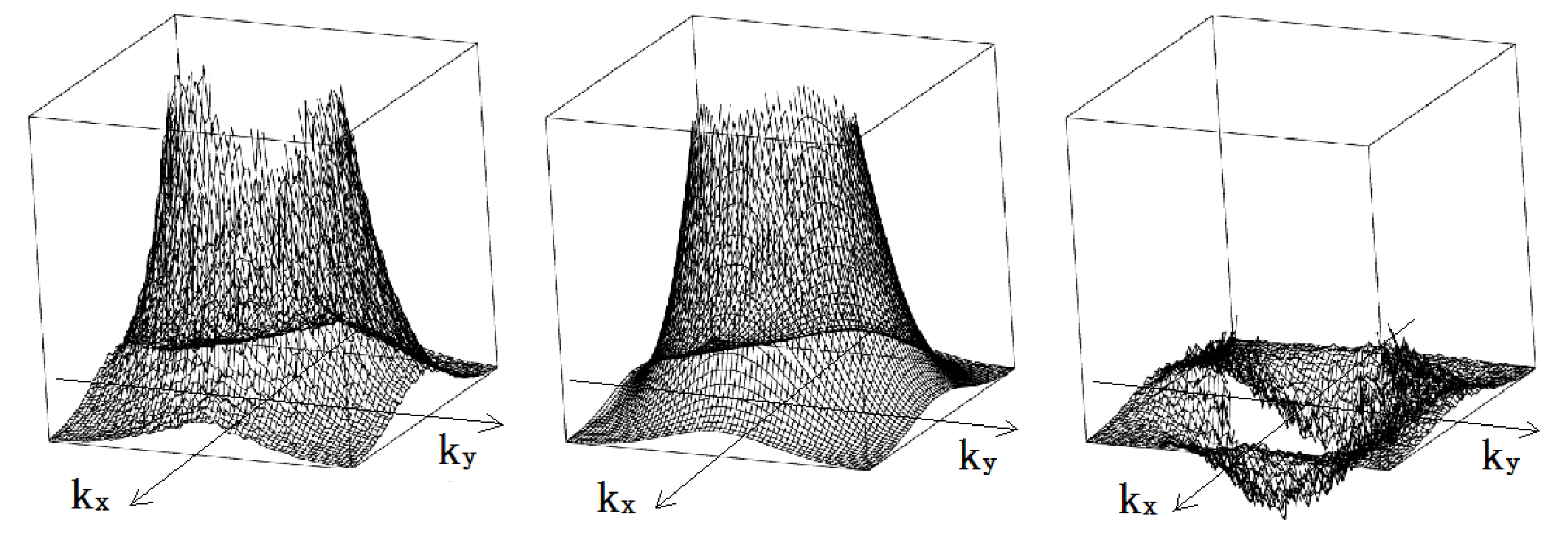}}
\caption{The shape of the PS of the source Poisson noise. The left figure is the stacked PS from the COSMOS data according to eq.(\ref{measure_C}) with sources of MAG\_AUTO $\ge$ 26. The middle one is $\left\vert\tilde{C}(\vec{k})\right\vert^2$ calculated from the COSMOS MultiDrizzle parameters, and scaled to best-fit the stacked PS of the COSMOS data. The right plot shows the differences between the left and the middle plots. All three plots have the same linear vertical scales.}
\label{poisson_comparison_cosmos}
\end{figure}

In fig.\ref{poisson_comparison_cosmos}, we show how well the calculated $\left\vert\tilde{C}(\vec{k})\right\vert^2$ from the COSMOS MultiDrizzle parameters can fit the stacked PS from real COSMOS images. The left plot is the stacked PS using about $4.3\times 10^5$ source-noise image pairs from the COSMOS data; the middle plot is the calculated $\left\vert\tilde{C}(\vec{k})\right\vert^2$, scaled to best-fit the left plot; the right plot shows the differences between the previous two plots. All three plots have the same linear verticle scales, and the bottom planes correspond to zero amplitude. Like fig.\ref{poisson_power_sdss}, the four sides on the bottom of each plot are the directions of $k_x$ and $k_y$ respectively, and the central region of each cube corresponds to small wave-numbers. The figure clearly shows that the shape of the PS at large wave-numbers can indeed be predicted/described well by the dithering and MultiDrizzle parameters. 

\begin{figure}
\centerline{\epsfxsize=9cm\epsffile{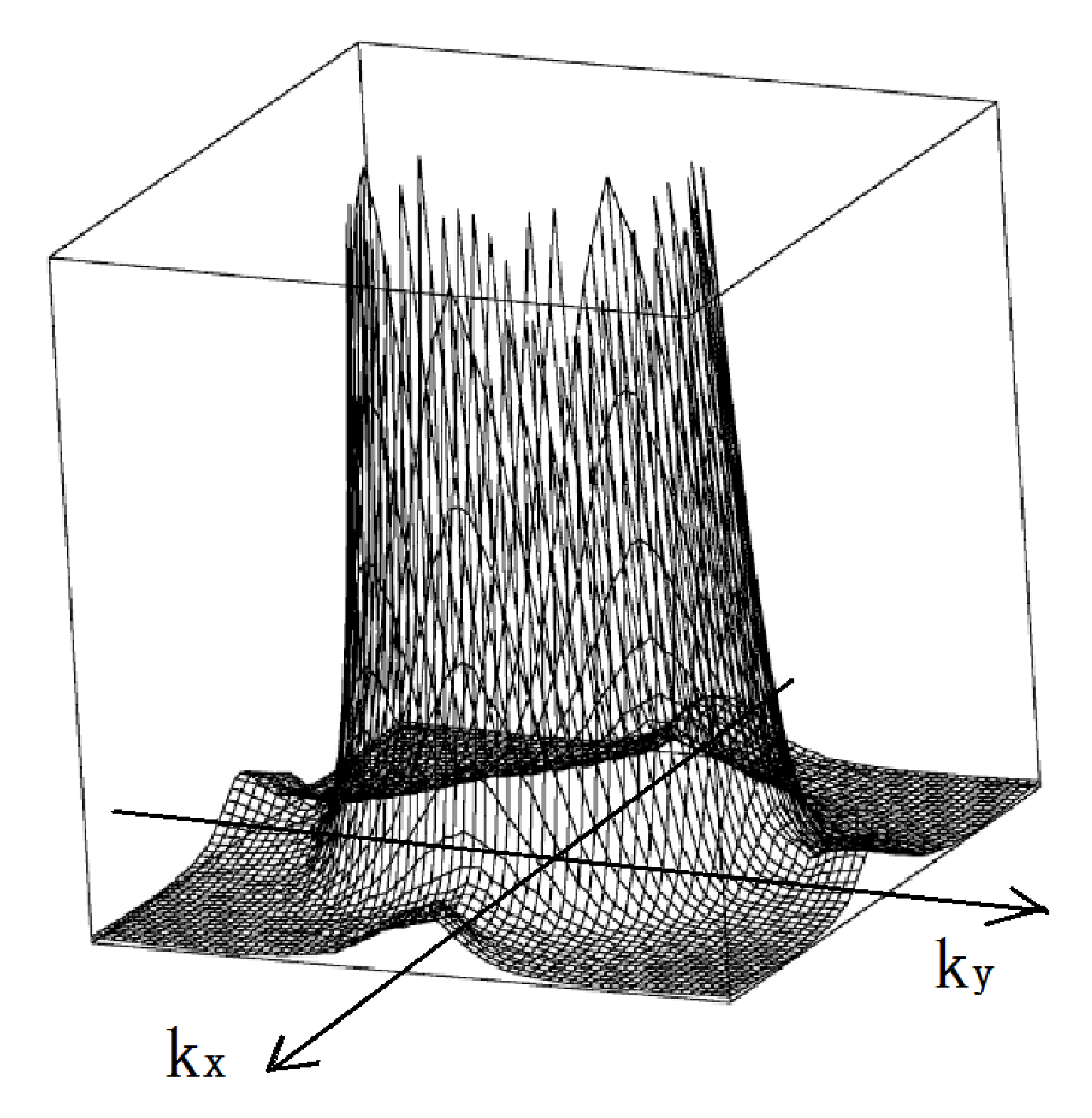}}
\caption{The stacked PS from sources of MAG\_AUTO between 22 and 23 in the COSMOS data.}
\label{power22}
\end{figure}

The stacked PS of brighter sources reveals more details of image properties. For example, fig.\ref{power22} shows the stacked PS from sources of MAG\_AUTO between 22 and 23. In this case, the source power is so large that the Poisson noise power is not quite visible even at large wave-numbers in the plot. We find that the source PS exhibits a spiky feature along the $k_x=0$ and $k_y=0$ directions, extending all the way to large wave-numbers. 

\begin{figure}
\centerline{\epsfxsize=9cm\epsffile{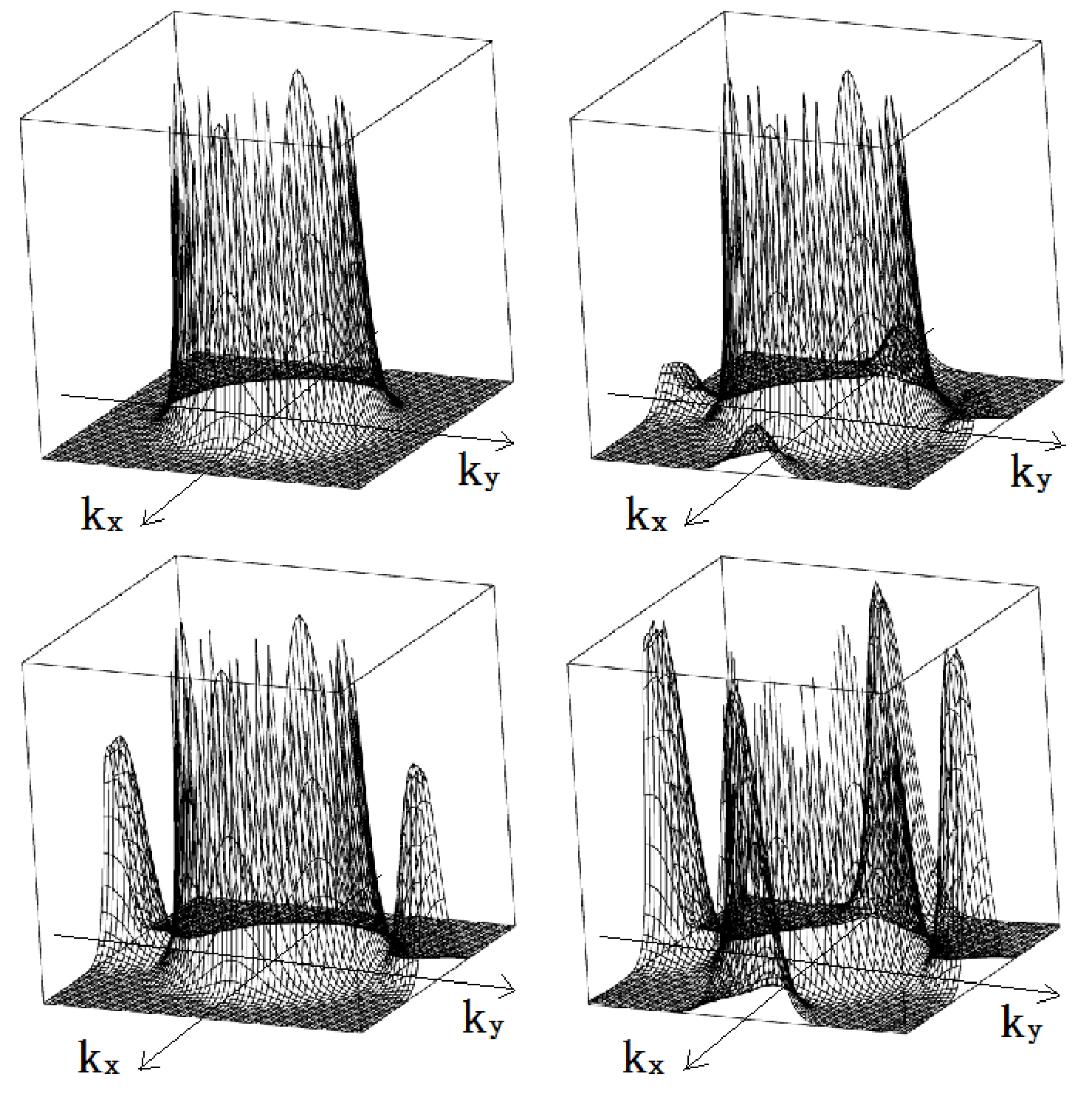}}
\caption{The PS of source image under four different weighting choices in MultiDrizzle: 1. (upper-left) all four exposures are weighted equally; 2. (upper-right) one exposure is down-weighted by 30\%; 3. (lower-left) two exposures (shifted along one side of the input pixels) are both down-weighted by 30\%; 4. (lower-right) one exposure has zero weight.}
\label{dither_weight_defects}
\end{figure}

This feature is possibly due to two reasons: 1. the spikes of the PSF due to the supporting rods of the telescope; 2. the unequal weightings of different exposures. In either case, the PS of this feature is proportional to the square of the source flux, unlike that of the source Poisson noise. The second factor can be simulated by changing the weights of the four exposures in the MultiDrizzle pipeline of our own. In fig.\ref{dither_weight_defects}, we show the source PS under four different weighting choices without Poisson noise or background noise: 1. (upper-left) all four exposures are weighted equally; 2. (upper-right) one exposure is down-weighted by 30\%; 3. (lower-left) two exposures (shifted along one side of the input pixels) are both down-weighted by 30\%; 4. (lower-right) one exposure has zero weight. 

\begin{figure}
\centerline{\epsfxsize=9cm\epsffile{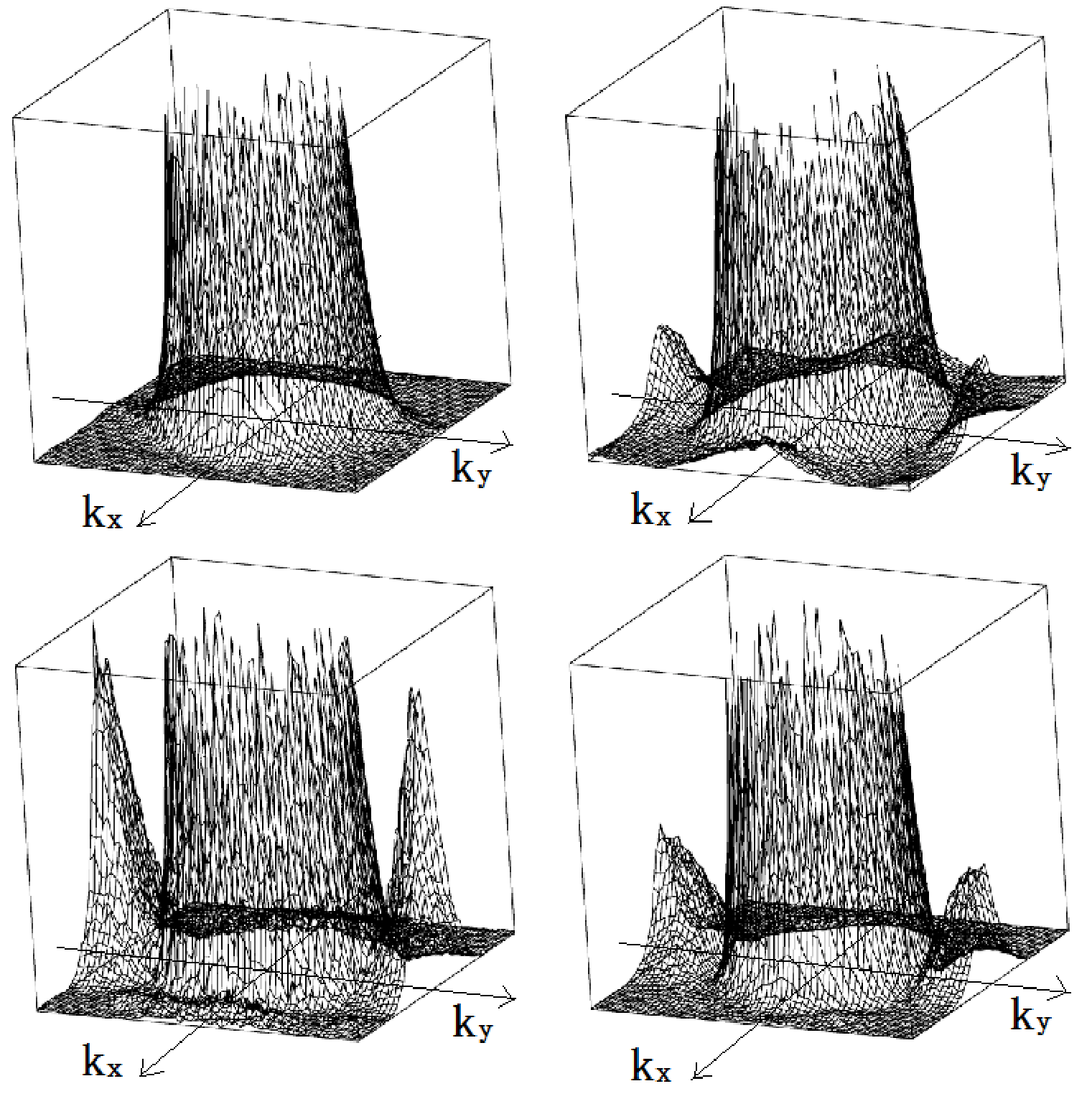}}
\caption{The PS of four source images that are randomly selected from COSMOS data with MAG\_AUTO between 20 and 22.}
\label{COSMOS_eg}
\end{figure}

It is clear that uneven weighting of the exposures in MultiDrizzle can lead to features of the source PS at large wave-numbers, which are not at all related to the Poisson noise. We indeed observe these features for individual source images in the COSMOS data, such as those shown in fig.\ref{COSMOS_eg}. In this case, to estimate the amplitude of the Poisson noise PS, we could avoid the affected regions in Fourier space by modifying eq.(\ref{shear_estimator_dis_para6}) as:
\begin{eqnarray}
\label{shear_estimator_dis_para7}
&&F^S=\frac{\sum_{\vec{k}_j\in S}\left\vert\tilde{f}^S(\vec{k}_j)\right\vert^2}{\sum_{\vec{k}_j\in S}\left\vert\tilde{C}(\vec{k}_j)\right\vert^2},\;\;\; F^B=\frac{\sum_{\vec{k}_j\in S}\left\vert\tilde{f}^B(\vec{k}_j)\right\vert^2}{\sum_{\vec{k}_j\in S}\left\vert\tilde{C}(\vec{k}_j)\right\vert^2}\\ \nonumber
&&S=\left\{\vec{k}\left\vert\vert\vec{k}\vert>k_c,\vert(\vec{k})_y\vert>k_{yc},\vert(\vec{k})_x\vert>k_{xc}\right.\right\}
\end{eqnarray}
in which $S$ is the set of wave vectors, which requires not only that the amplitude of $\vert\vec{k}\vert$ is larger than $k_c$, but its two components along the $k_x$ and $k_y$ directions should also have amplitudes larger than given values $k_{xc}$ and $k_{yc}$ respectively. $k_c$, $k_{xc}$, and $k_{yc}$ should all be chosen appropriately to reflect the nature of the images. 

\begin{figure}
\centerline{\epsfxsize=9cm\epsffile{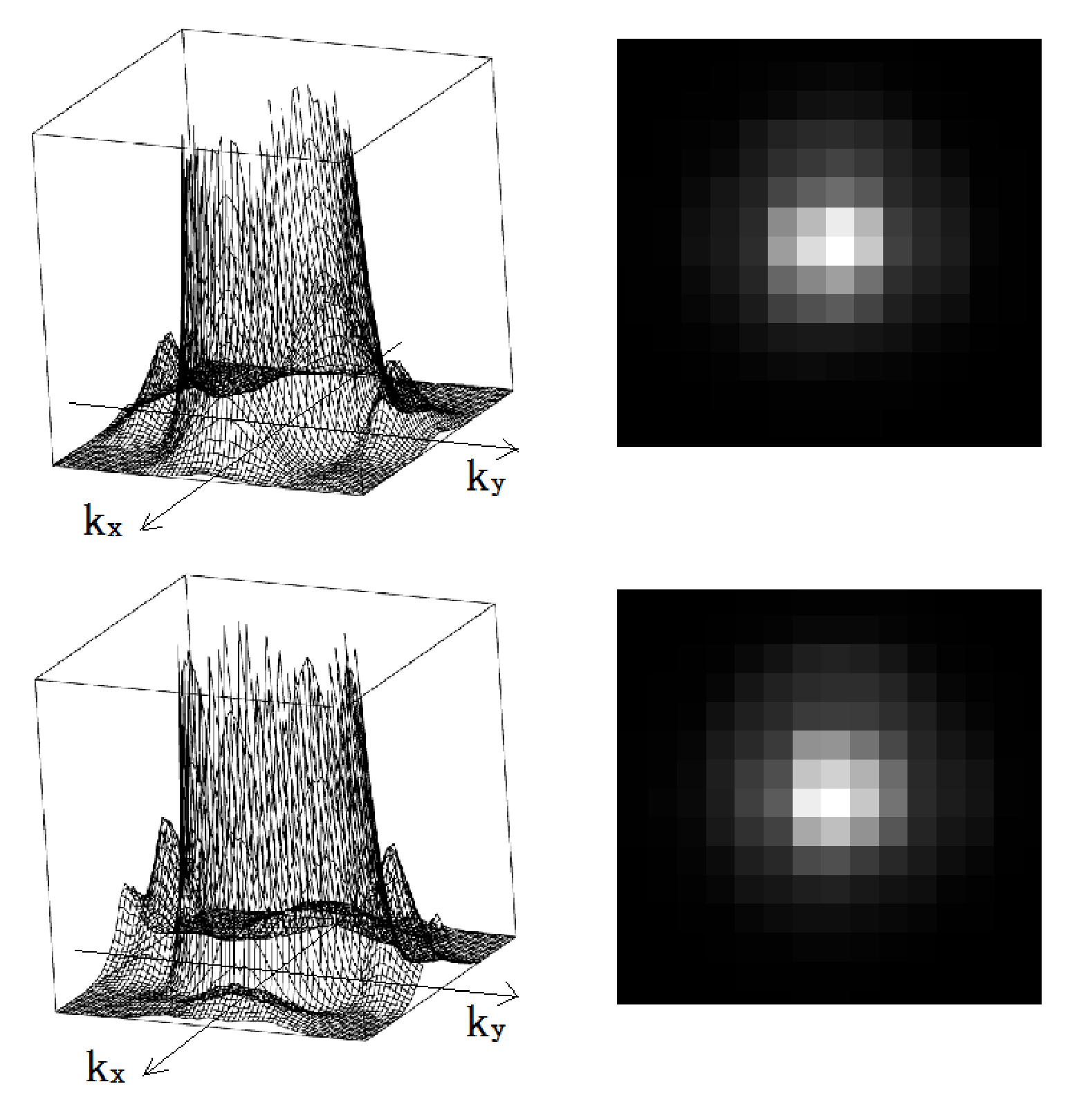}}
\caption{The PS (on the left side) of two source images (on the right side) found in the COSMOS data. Their PS exhibit features at large wave-numbers that cannot be easily explained by uneven weightings of the exposures. }
\label{anomaly}
\end{figure}

Note that the definition of the set $S$ could be quite flexible and general, as long as within the region defined by $S$, the PS of the image is dominated by the Poisson noise. This fact enables us to easily adapt our method to remove the source Poisson noise effect for images with special features at large wave-numbers in Fourier space. For example, fig.\ref{anomaly} shows the Fourier transformations (on the left side of the figure) of two source images (on the right side) found in the COSMOS data, whose spectral shapes exhibit features at large wave-numbers that cannot be easily explained by uneven weightings of the exposures. It is also useful to note that such features/anomalies can be easily identified in Fourier space, but not in real space. A further study of this topic will require more details of the data reduction pipeline and observing conditions.      

We caution that our test of shear recovery accuracy reported so far are still about images generated through simple pipelines, \eg, equal weightings for all four exposures in MultiDrizzle for every image. We find that uneven weightings of the exposures could lead to percent-level systematic shear errors in our method given the observing condition of COSMOS/ACS. In this sense, even weighting is strongly preferred. This is likely achievable by increasing the number of exposures. A further study of this issue requires a good understanding of the image data reduction pipeline, which will be the focus of a future work.  

\begin{figure}
\centerline{\epsfxsize=9cm\epsffile{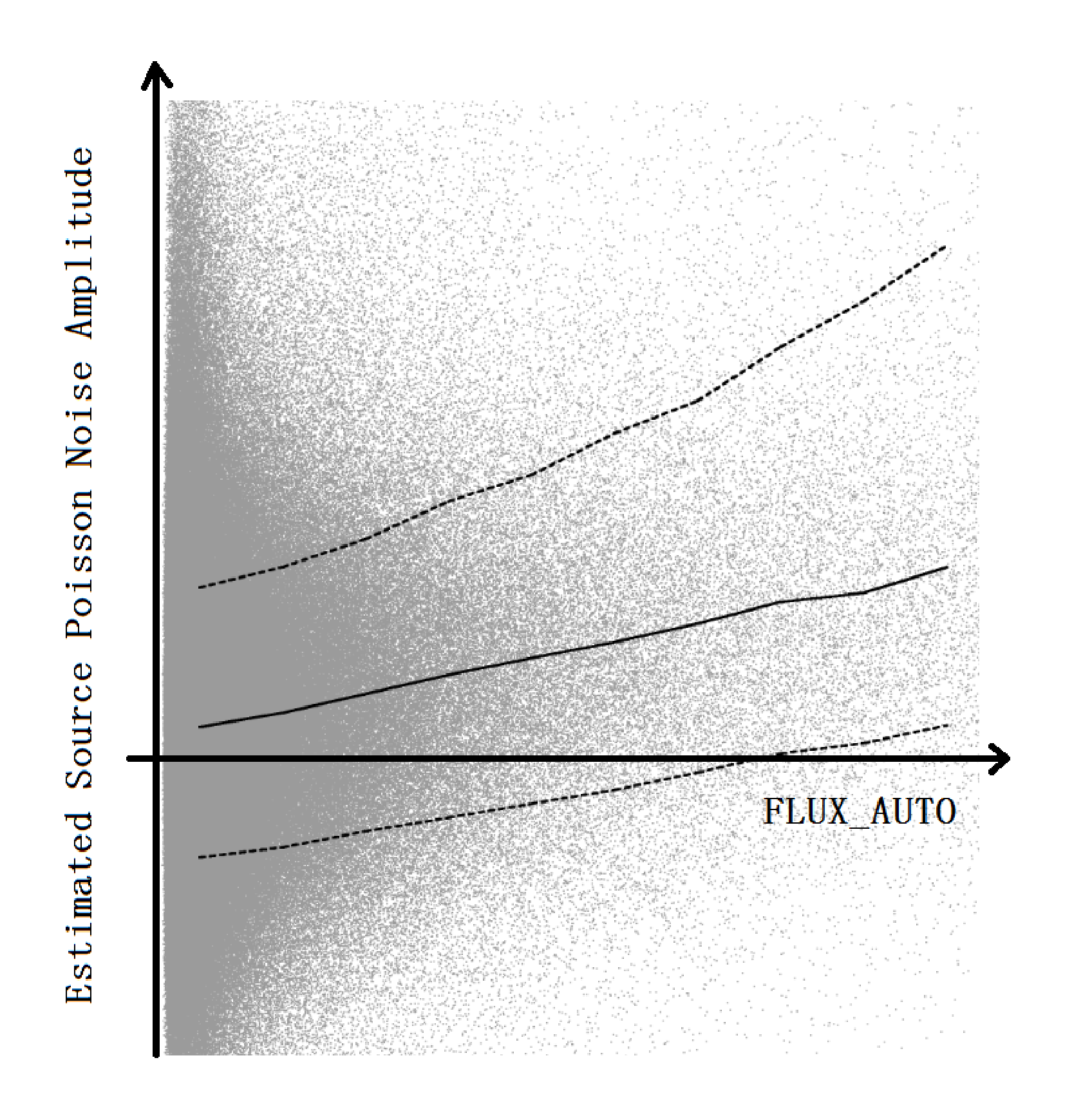}}
\caption{The relation between the PS of the source Poisson noise evaluated using $F^S-F^B$ defined in eq.(\ref{shear_estimator_dis_para7}) and the FLUX\_AUTO of the source. The result from each source is presented as a grey dot in the figure. The data is grouped into ten bins equally spaced in FLUX\_AUTO. The solid curve connects the medians in the neighboring bins, and 68 percent of the points are within the region enclosed by the dotted lines in each bin. Both axes are in linear scales.}
\label{power_flux_relation}
\end{figure}

Finally, as a consistency check, we can plot $F^S-F^B$ defined in eq.(\ref{shear_estimator_dis_para7}) against the source flux FLUX\_AUTO for each image. We shall expect a linear relation between these two quantities according to the nature of Poisson noise, similar to what is shown in fig.\ref{power_flux_sdss}. This test is done by setting $k_c$ to 0.75 times the Nyquist spatial-frequency along either side of the image, and $k_{xc}$ and $k_{yc}$ to 3 times the frequency resolution along either direction. The results are shown in fig.\ref{power_flux_relation}. In the figure, both axes are in linear scales, and the joint of the two axes corresponds to zero FLUX\_AUTO and zero value of the PS of the source Poisson noise. We include the data from sources of MAG\_AUTO larger than 23, each of which is shown as a grey dot in the figure. We group the data into ten bins equally spaced in FLUX\_AUTO. The solid curve connects the median amplitudes of the source Poisson noise in neighboring bins, and 68 percent of the points are within the region enclosed by the dotted lines in each bin. The figure confirms the linear relation between the PS of the source Poisson noise and the source flux. The scatter is mostly due to the background noise. Note that towards the right end of the figure, \ie, for bright sources, the upper part of the 68-percent contour curves upward, implying overestimates of the Poisson noise power. This phenomenon is consistent with our finding that the source power could be present at large wave numbers due to, \eg, uneven weightings of the exposures in MultiDrizzle. Large values of $k_{xc}$ and $k_{yc}$ may better remove this problem.

\section{Conclusions}
\label{conclusions}

The majority of galaxies identified in a galaxy survey are at the faint end of the luminosity function. They therefore contribute the dominant amount of shape information in weak lensing measurements. On the other hand, faint galaxies have relatively large photon counting noises, \ie, the source Poisson noise, which can lead to significant bias in shear recovery as shown in Z11. This is an important issue not only for space-based weak lensing surveys (\eg, COSMOS, Euclid), in which the low background level allows very faint galaxies to be identified as valid sources, but also for ground-based projects (\eg, CFHTLenS, DES, LSST) if images of short exposure times are used for shear measurement. The purpose of this work is to remove the shear recovery errors due to the source Poisson noise by extending the shear measurement method of Z11.   

The key idea is to realize that for images made of single exposures, the average PS of the Poisson noise is independent of the wave-number in Fourier space, while the PS of the signal that are useful for shear measurement mostly concentrate at the low wave-number regions because of the PSF. In other words, Fourier transformation provides a natural way of separating the source signal from its Poisson noise. This fact has led to a number of important techniques in image processing, including, \eg, Wiener filtering. In the method of Z11, since the components of the shear estimators are linear functions of the source PS, we can directly remove the Poisson noise PS that is estimated at large wave-numbers. This idea has led to a modified version of Z11 defined in eq.(\ref{shear_measure},\ref{shear_estimator_dis},\ref{shear_estimator_dis_para1},\ref{shear_estimator_dis_para3},\ref{shear_estimator_dis_para4}), which can indeed successfully remove the shear measurement bias due to the source Poisson noise, as shown with numerical examples in \S\ref{ne}.

In space-based weak lensing measurements, dithering is often required for increasing the sampling of galaxy image. In this case, each source image is usually combined from a number of exposures that are taken at positions shifted by sub-pixel scales. The popular combination technique is called MultiDrizzle, in which the value of each output pixel is a weighted sum of those of the neighboring input pixels. This type of image processing introduces correlations between the Poisson noise on neighboring output image pixels. Consequently, the Poisson noise PS becomes dependent on the wave-number. In this case, we show in \S\ref{dithering} that the spectral shape of the Poisson noise in Fourier space is determined by the parameters in dithering and MultiDrizzle. Similar to what is done to images of single exposures, we can estimate the amplitude of the Poisson noise PS at large wave-numbers, and then subtract its contamination on all scales according to the pre-determined spectral shape. Correspondingly, the forms of the shear estimators addressing this type of treatment are given in eq.(\ref{shear_measure},\ref{shear_estimator_dis},\ref{shear_estimator_dis_para1},\ref{shear_estimator_dis_para5},\ref{shear_estimator_dis_para6}). We have demonstrated in \S\ref{ntest2} that for images processed by MultiDrizzle, shear recovery can still achieve a sub-percent level accuracy with the modified Z11. 

With the SDSS data, we have shown that the Poisson noise PS of single-exposure images are indeed independent of the wave number $k$. Using the dithering and MultiDrizzle parameters of the COSMOS/ACS data, we have predicted the non-trivial shape of the Poisson noise PS, and verified that at least at large wave-numbers, the prediction agrees very well with the shape of the stacked PS of real COSMOS sources with MAG\_AUTO larger than 26, as shown in fig.\ref{poisson_comparison_cosmos}. 

The COSMOS data further shows that the source PS sometimes contains features at large wave-numbers, \eg, along the $k_x=0$ and $k_y=0$ directions, that are not due to the Poisson noise. We argue that this phenomenon could be due to two reasons: 1. the spikes of the PSF due to the supporting rods of the telescope; 2. uneven weighting of the exposures. We provide strong support for the second point in fig.\ref{dither_weight_defects}. This problem could however be easily remedied by further generalizing the way of estimating the Poisson noise amplitude in eq.(\ref{shear_estimator_dis_para7}), \ie, by more carefully choosing the regions in Fourier space that are dominated by the Poisson noise power. The details may rely on the quality of the imaging data and the ways of combining the multiple exposures into single images. With this effect being taken into account, we have shown in fig.\ref{power_flux_relation} that the PS of the source Poisson noise in the COSMOS/ACS data is proportional to the source flux, agreeing with what is expected. To conclude, we find that to remove the source Poisson noise effect on shear measurement in general, the shear estimators of Z11 could be extended to forms defined by eq.(\ref{shear_measure},\ref{shear_estimator_dis},\ref{shear_estimator_dis_para1},\ref{shear_estimator_dis_para5},\ref{shear_estimator_dis_para7}), with set $S$ in eq.(\ref{shear_estimator_dis_para7}) being appropriately chosen to avoid possible features of images at large wave-numbers.

It is useful to note that for shear measurement methods based on model-fitting the galaxy morphology with a given set of parameters, the noise bias is typically hard to avoid due to the nonlinear dependence of shear estimators on the image surface brightness distribution \citep{refregier12,mv12,miller13,vio14}. This problem is not present in our method, since each component of our shear estimators has a linear dependence on the image power spectrum, and can be removed unambiguously. Note that this fact may lead the reader to think that our shear measurement method is a image-stacking method, in which the shear signal are derived from many stacked images \citep{lewis09}. This is simply not true, as our shear estimator is defined on each source image. 

Our shear measurement method is also significantly different from other methods defined in Fourier space. For example, the FDNT method of \citet{bernstein10} relies on the null testing of the moments of the Fourier modes of the source image to recover shear. It however still suffers from the shear bias due to the photon noise and the ellipticity gradients of the image. In the Bayesian Fourier domain (BFD) method proposed by \citet{ba14}, it is not clear how to treat the noise bias. Furthermore, the BFD method requires the unlensed distribution of galaxy moments as a prior, which is not necessary in our method.

To fully achieve the sub-percent level accuracy in shear measurements for the ongoing and up-coming large scale galaxy surveys, there are a number of other issues to be concerned at different levels. For example, the reconstruction of the PSF's at the positions of the galaxies using the stellar images is a well known unsolved problem in this field. Immediately related to this work, we need to understand better that in MultiDrizzle, if it is feasible to give the same weightings to all the participating exposures, which is a preferred scheme in the shear measurement method of this paper. This problem is in principle solvable by increasing the number of exposures, which also provides more options for cosmic ray rejections. Regarding image qualities, we still have a number of issues that may be addressed more carefully, including the image selection criteria, charge transfer efficiency \citep{massey10,rhodes10,massey10b}, pixel defects, differential telescope distortion, and some details in the image reduction pipeline. Some of these issues should be studied in a more specific framework, \ie, for a given set of instruments and observing conditions. This is an important direction of our future work. 

Finally, it is perhaps not difficult to find out that the current forms of our shear estimators tend to weight the bright/large galaxies more than the faint/small ones, which is certainly not the optimal weighting scheme. In our next work, we will explore ways to weight the shear estimator components from each galaxy, so that the statistical error of shear measurement can be minimized without incurring systematic errors.

\acknowledgments{The authors thank Jason Rhodes and Richard Massey for help on the COSMOS/ACS data, and Longlong Feng, Lei Hao, Nobuhiko Katayama, Eiichiro Komatsu, Alexie Leauthaud, Chengze Liu, Yi Mao, Pengjie Zhang for useful discussions, Gary Bernstein, and our referees for comments on an early version of this paper. JZ is supported by the national science foundation of China (Grant No. 11273018, 11433001), the national basic research program of China (Grant No. 2013CB834900, 2015CB857001), the national “Thousand Talents Program” for distinguished young scholars, a grant(No.11DZ2260700) from the Office of Science and Technology in Shanghai Municipal Government, and the T.D. Lee Scholarship from the High Energy Physics Center of Peking University. JZ is previously supported by the TCC Fellowship of Texas Cosmology Center of the University of Texas at Austin, where a part of this work was done. Some numerical calculations of this work was done using the High Performance Computing (HPC) resources of the Texas Advanced Computing Center.}


\label{lastpage}
\end{document}